\documentclass[twocolumn,showpacs,superscriptaddress,prb,aps,floatfix]{revtex4-2}

\usepackage{graphicx}
\usepackage{rotating}
\usepackage{amsmath}
\usepackage{float}
\usepackage{rotating}
\usepackage{booktabs}
\usepackage{multirow}
\usepackage{supertabular}
\usepackage[T1]{fontenc}
\usepackage{upgreek}
\usepackage{comment}
\usepackage{amssymb}
\usepackage{enumitem} 
\usepackage{siunitx}
\usepackage{caption}
\usepackage{booktabs}
\usepackage{placeins}
\usepackage{xcolor}
\usepackage{makecell}
\usepackage{xr}
\usepackage{url}
\definecolor{darkteal}{HTML}{254949}
\newcommand{\ignore}[1]{}
\hfuzz=\maxdimen
\def\tsc#1{\csdef{#1}{\textsc{\lowercase{#1}}\xspace}}
\tsc{WGM}
\tsc{QE}
\tsc{EP}
\tsc{PMS}
\tsc{BEC}
\tsc{DE}

\begin{document}

\title{Accelerated Discovery of Semiconductors with Ultralow Lattice Thermal Conductivity Using Domain Knowledge and uMLP}  



%








\author{Yateng Wang}
\affiliation{Key Laboratory of Advanced Materials and Devices for Post-Moore Chips, Ministry of Education, Beijing Key Laboratory for Magneto-Photoelectrical Composite and Interface Science, School of Mathematics and Physics, University of Science and Technology Beijing, Beijing 100083, People's Republic of China}

\author{Chenxi Li}
\affiliation{MOE Key Laboratory of Materials Physics and Chemistry under Extraordinary Conditions \& Shaanxi Provincial Key Laboratory of Condensed Matter Structure and Properties, School of Physical Science and Technology, Northwestern Polytechnical University, Xi'an, 710072, People's Republic of China}

\author{Jiabin Li}
\affiliation{Key Laboratory of Advanced Materials and Devices for Post-Moore Chips, Ministry of Education, Beijing Key Laboratory for Magneto-Photoelectrical Composite and Interface Science, School of Mathematics and Physics, University of Science and Technology Beijing, Beijing 100083, People's Republic of China}

\author{Zhonghao Xia}
\affiliation{Key Laboratory of Advanced Materials and Devices for Post-Moore Chips, Ministry of Education, Beijing Key Laboratory for Magneto-Photoelectrical Composite and Interface Science, School of Mathematics and Physics, University of Science and Technology Beijing, Beijing 100083, People's Republic of China}

\author{Lingzhi Cao}
\affiliation{Key Laboratory of Advanced Materials and Devices for Post-Moore Chips, Ministry of Education, Beijing Key Laboratory for Magneto-Photoelectrical Composite and Interface Science, School of Mathematics and Physics, University of Science and Technology Beijing, Beijing 100083, People's Republic of China}

\author{Jianye Liu}
\affiliation{Key Laboratory of Advanced Materials and Devices for Post-Moore Chips, Ministry of Education, Beijing Key Laboratory for Magneto-Photoelectrical Composite and Interface Science, School of Mathematics and Physics, University of Science and Technology Beijing, Beijing 100083, People's Republic of China}

\author{Jinming Zhai}
\affiliation{Key Laboratory of Advanced Materials and Devices for Post-Moore Chips, Ministry of Education, Beijing Key Laboratory for Magneto-Photoelectrical Composite and Interface Science, School of Mathematics and Physics, University of Science and Technology Beijing, Beijing 100083, People's Republic of China}

\author{Xingchen Shen}
\email{xingchen.shen@nwpu.edu.cn}
\affiliation{MOE Key Laboratory of Materials Physics and Chemistry under Extraordinary Conditions \& Shaanxi Provincial Key Laboratory of Condensed Matter Structure and Properties, School of Physical Science and Technology, Northwestern Polytechnical University, Xi'an, 710072, People's Republic of China}

\author{Jiangang He}
\email{jghe2021@ustb.edu.cn}
\affiliation{Key Laboratory of Advanced Materials and Devices for Post-Moore Chips, Ministry of Education, Beijing Key Laboratory for Magneto-Photoelectrical Composite and Interface Science, School of Mathematics and Physics, University of Science and Technology Beijing, Beijing 100083, People's Republic of China}

\date{\today}












\begin{abstract}
While lattice thermal conductivity ($\kappa_{\mathrm{L}}$) is one of the fundamental properties of solids, the high computational cost of anharmonic phonon scattering limits the search for materials with exceptionally low or high $\kappa_{\mathrm{L}}$. In this work, we identify semiconductors with exceptionally low or high $\kappa_{\mathrm{L}}$ by combining a universal machine-learning interatomic potential with two harmonic descriptors: the average atomic mean-square displacement (MSD) and the lowest zone-center optical phonon frequency ($\omega_{\min}^{\mathrm{opt}}$). We fine-tuned the universal machine-learning interatomic potential MatterSim on second-order interatomic force constants from PhononDB and used it to calculate MSD and $\omega_{\min}^{\mathrm{opt}}$ in a large scale. Of 8,974 compounds screened in the Open Quantum Materials Database, these descriptors identified 463 low-$\kappa_{\mathrm{L}}$ and 49 high-$\kappa_{\mathrm{L}}$ candidates. First-principles transport calculations at 300~K predicted $\kappa_{\mathrm{L}}$ < 1 Wm$^{-1}$K$^{-1}$ for 104 of the 149 low-conductivity candidates evaluated and $\kappa_{\mathrm{L}}$ > 100 Wm$^{-1}$K$^{-1}$ for 15 of the 34 high-conductivity candidates evaluated. By tracing both conductivity extremes to harmonic signatures of phonon propagation and scattering phase space, this work provides a common physical basis for discovering semiconductors that either suppress or efficiently conduct heat across a broad range of crystal chemistries.
\end{abstract}




 

\maketitle

\section{Introduction}\label{}
Lattice thermal conductivity ($\kappa_{\mathrm{L}}$) is a key determinant of heat transport in solids and is therefore important in a broad range of technological applications. Semiconductors with ultralow $\kappa_{\mathrm{L}}$ are desirable for thermal insulation and thermoelectric energy conversion~\cite{Thakare2021,MEHTA20201336,CHANG201850,Ghosh2022}, whereas materials with high $\kappa_{\mathrm{L}}$ are essential for efficient thermal management and heat dissipation in high-power electronic devices~\cite{Hu2020,Qian2021}. These contrasting technological demands have made the discovery and rational design of materials with extreme $\kappa_{\mathrm{L}}$ an important frontier in contemporary materials research.

Within a simple kinetic-theory framework~\cite{Yang2004}, $\kappa_\mathrm{L} = \frac{1}{3} C \nu_\mathrm{g}^2 \tau$, where $C$, $\nu_\mathrm{g}$, and $\tau$ denote the heat capacity, phonon group velocity, and phonon relaxation time, respectively~\cite{https://doi.org/10.1002/andp.19293950803}. These three quantities jointly determine $\kappa_{\mathrm{L}}$ within this framework. Extensive studies have identified a variety of mechanisms that strongly influence phonon transport, including complex crystal structures~\cite{Jia-JunSHEN260,liu2022efficient}, rattling atoms~\cite{PhysRevLett.114.095501,doi:10.1126/science.aar8072,https://doi.org/10.1002/anie.201605015,christensen2008avoided}, bonding heterogeneity~\cite{pal2018bonding}, occupied antibonding states~\cite{10.1002/adfm.202108532,10.1002/advs.202417292}, lone electron pairs~\cite{C2EE23391F,PhysRevLett.107.235901,maria2023metavalent}, and resonant bonding~\cite{ji2022functional,YUE201889}. In general, materials with strong lattice anharmonicity~\cite{li2015orbitally,PhysRevLett.117.075502}, heavy constituent elements~\cite{shi2018new}, and weak chemical bonds~\cite{10.1002/adfm.202108532} tend to exhibit low $\kappa_{\mathrm{L}}$. By contrast, high $\nu_{\mathrm{g}}$ and weak anharmonic phonon scattering are essential for ultrahigh $\kappa_{\mathrm{L}}$~\cite{SLACK1973321}. Light constituent atoms, fourfold coordination~\cite{PhysRevX.10.041029}, and strong chemical bonding favor high-frequency, highly dispersive phonon modes and thus large $\nu_{\mathrm{g}}$. Simple crystal structures and weak lattice anharmonicity, in turn, help suppress phonon--phonon scattering and sustain long phonon lifetimes~\cite{PhysRevB.88.214303,PhysRevB.80.125203}. These observations demonstrate that $\kappa_{\mathrm{L}}$ reflects a complex interplay among structural, chemical, and dynamical factors, making the identification of materials with exceptionally low or high $\kappa_{\mathrm{L}}$ challenging but important.

The discovery of materials with exceptionally low or high $\kappa_{\mathrm{L}}$ has traditionally relied on experimental trial and error. With advances in computational capabilities, lattice heat-transport theory, and numerical algorithms~\cite{Hooton01041955,LI20141747,Simoncelli2019,Togo_2023,PhysRevLett.113.185501}, it has become feasible to calculate the $\kappa_{\mathrm{L}}$ of inorganic materials entirely at the DFT level. Ojih et al. calculated $\kappa_{\mathrm{L}}$ for 8077 inorganic compounds within the three-phonon scattering framework~\cite{D3TA04874H}. Ohnishi et al. constructed a DFT-based database of $\kappa_{\mathrm{L}}$ for 6500 inorganic compounds~\cite{WOS:001739467100001}, incorporating three-phonon scattering as well as coherent contributions. Li et al.~\cite{Li2025} performed high-throughput $\kappa_{\mathrm{L}}$ calculations for 773 cubic and tetragonal crystals spanning diverse chemistries and structures, accounting for self-consistent phonon renormalization~\cite{Hooton01041955,PhysRevLett.17.89}, three- and four-phonon scattering~\cite{PhysRevB.96.161201}, and off-diagonal heat flux~\cite{Simoncelli2019}. However, accurate $\kappa_{\mathrm{L}}$ calculations based on anharmonic lattice dynamics and the phonon Boltzmann transport equation require higher-order interatomic force constants derived from DFT. These calculations are computationally demanding, severely limiting large-scale screening across broad chemical spaces and diverse crystal structures. To address this limitation, high-throughput computational and data-driven approaches have increasingly been developed to accelerate the search for materials with extreme $\kappa_{\mathrm{L}}$~\cite{PhysRevX.4.011019,Pal2022}.

For low-$\kappa_{\mathrm{L}}$ materials, one strategy is to perform high-throughput calculations on compounds pre-screened using physically interpretable descriptors derived from crystal structure, chemical bonding, and lattice dynamics. Some of the authors of this work developed an efficient screening strategy based on local coordination environments and weak chemical bonding, which led to the identification of 30 previously unexplored low-$\kappa_{\mathrm{L}}$ compounds spanning 17 structural prototypes~\cite{10.1002/adfm.202108532}. Shen et al. introduced a universal descriptor based on atomic number and $\nu_{\mathrm{g}}$ to screen more than 25,000 crystalline materials. They identified CsAg$_2$I$_3$, which exhibits a glass-like ultralow $\kappa_{\mathrm{L}}$ of 0.15--0.16 Wm$^{-1}$K$^{-1}$ between 170 and 400 K; this behavior was attributed to its intrinsically low $\nu_{\mathrm{g}}$, strong phonon anharmonicity, and structural complexity~\cite{Shen2025}. Another strategy uses machine-learning models to predict $\kappa_{\mathrm{L}}$ directly or to predict closely related quantities. Seko et al. used Bayesian optimization, initialized with first-principles anharmonic lattice-dynamics calculations for 101 compounds, to search a library of 54,779 materials and identify 221 compounds with very low $\kappa_{\mathrm{L}}$~\cite{PhysRevLett.115.205901}. Wang et al. developed an XGBoost model based on compositional and crystal-structure descriptors to predict $\kappa_{\mathrm{L}}$ in crystalline materials, achieving a mean absolute error (MAE) of 2.13~Wm$^{-1}$K$^{-1}$. Applying the trained model to compounds in the Inorganic Crystal Structure Database (ICSD)~\cite{Zagorac:in5024}, they identified TlBiTe$_2$ and CsICl$_2$ as promising ultralow-$\kappa_{\mathrm{L}}$ materials, with $\kappa_{\mathrm{L}}$ values of 0.20 and 0.22 Wm$^{-1}$K$^{-1}$ at 300 K, respectively~\cite{10.1021/acs.jpcc.9b11610}. Anam et al. used the Elemental-SDNNFF deep-learning interatomic potential to screen thousands of cubic double perovskites and identified Cs$_2$HgPtCl$_6$ as having an ultralow $\kappa_{\mathrm{L}}$ of 0.071 Wm$^{-1}$K$^{-1}$ at room temperature. This low value was attributed to weak chemical bonding, strong anharmonicity, and reduced phonon transport from both diagonal and off-diagonal contributions~\cite{advs.202515766}.

For high-$\kappa_{\mathrm{L}}$ materials, both physically informed descriptors and machine-learning approaches have been utilized. Liu et al. used symmetry-operation counts and primitive-cell atom counts to screen 803 carbon allotropes for high $\kappa_{\mathrm{L}}$, identifying bcc-C$_6$ as a high-$\kappa_{\mathrm{L}}$ material (2198 Wm$^{-1}$K$^{-1}$ at room temperature)~\cite{10.1021/acsaem.2c02970}. Ju et al. developed a feature-based transfer-learning model that used the three-phonon scattering phase space ($\mathrm{P}_3$) as an auxiliary property and applied it to screen more than 60,000 crystalline compounds. Their search identified 14 promising candidates with high calculated $\kappa_{\mathrm{L}}$, including BAs, diamond, BN, mixed diamond--lonsdaleite carbon structures, and two BC$_2$N polymorphs~\cite{PhysRevMaterials.5.053801}. Guo et al. combined a deep generative model with structural symmetry and similarity descriptors, active learning, and machine-learning interatomic potentials to screen 100,000 generated carbon structures. They identified 34 carbon polymorphs with predicted room-temperature $\kappa_{\mathrm{L}}$ exceeding 800 Wm$^{-1}$K$^{-1}$ in at least one crystallographic direction~\cite{Guo2025}. These studies demonstrate that physically interpretable descriptors and machine-learning methods offer complementary strategies for accelerating the discovery of high-$\kappa_{\mathrm{L}}$ materials.

\begin{figure*}[tph!]
	\centering
	\includegraphics[clip,width=1.0\linewidth]{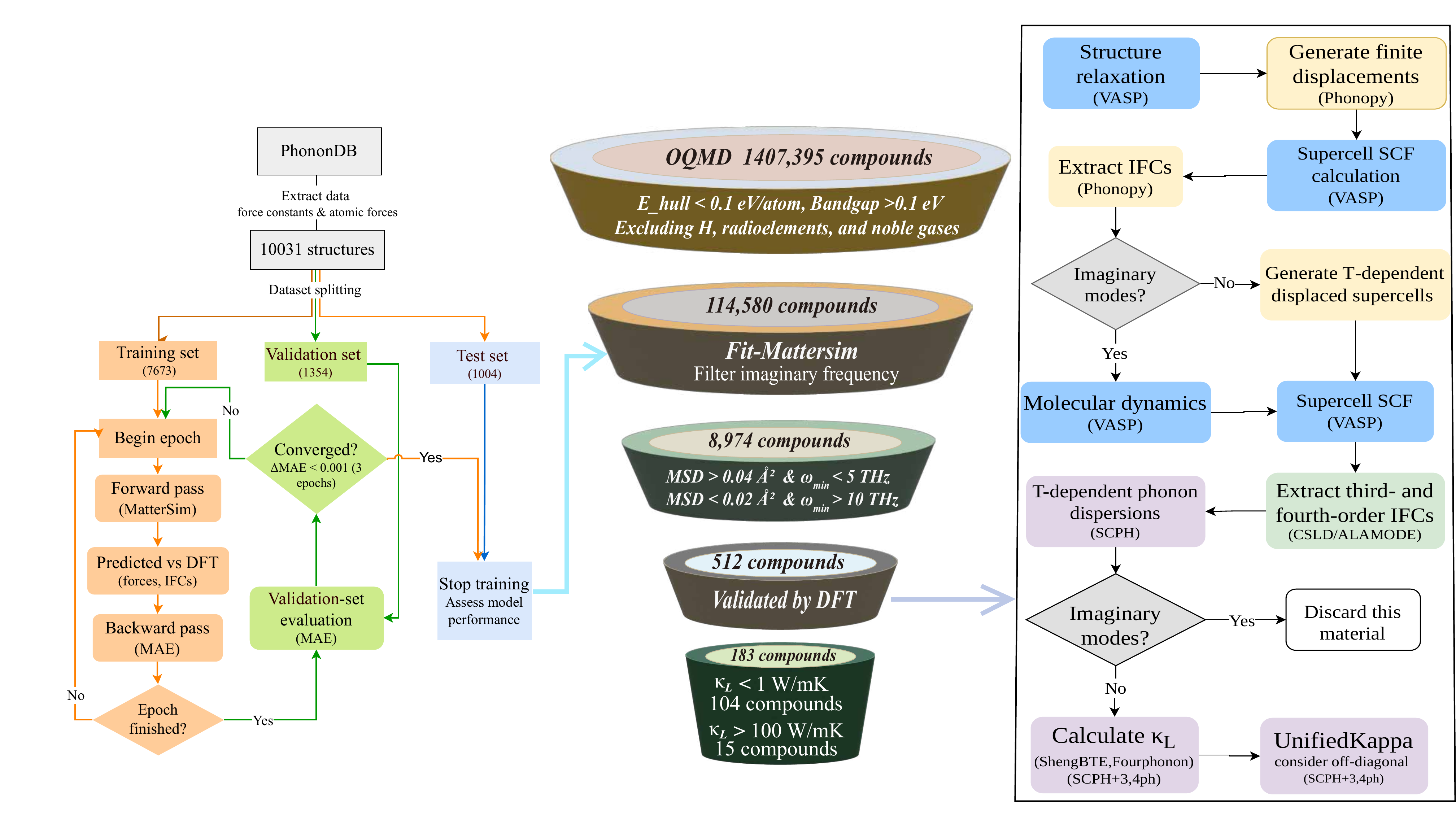}
	\caption{Overall workflow of this study. The left panel illustrates the fine-tuning procedure of the MatterSim model for accurate prediction of lattice thermal conductivity. The middle panel presents the hierarchical screening workflow adopted to efficiently identify candidate materials with extreme lattice thermal conductivity from the OQMD database. The right panel shows the first‑principles calculation pipeline for precise evaluation of lattice thermal conductivity, which validates the screened material candidates.}
	\label{fig:filter_step}
\end{figure*}

In this work, we develop an efficient framework for rapidly identifying materials with extreme $\kappa_{\mathrm{L}}$, as illustrated in Fig.~\ref{fig:filter_step}. To improve the accuracy of lattice-dynamical properties, we fine-tune the machine-learning interatomic potential MatterSim~\cite{yang2024mattersimdeeplearningatomistic} using second-order interatomic force constants derived from first-principles calculations. We then use the refined model to calculate the  mean square displacement (MSD) and the minimum optical phonon frequency ($\omega_{\min}^{\mathrm{opt}}$) for thermodynamically stable and metastable semiconducting compounds in the Open Quantum Materials Database (OQMD)~\cite{OQMD}. Based on these two descriptors, we select materials with MSD > 0.04 $\mathrm{\text{\AA}^2}$ and $\omega_{\min}^{\mathrm{opt}}$ below 5 THz as candidate low-$\kappa_{\mathrm{L}}$ materials. Materials with MSD < 0.02 $\mathrm{\text{\AA}^2}$ and $\omega_{\min}^{\mathrm{opt}}$ above 10 THz are identified as candidate high-$\kappa_{\mathrm{L}}$ materials. To evaluate the screening strategy, we further assess representative candidates using first-principles $\kappa_{\mathrm{L}}$ calculations. Among the 149 evaluated low-$\kappa_{\mathrm{L}}$ candidates, 104 exhibit $\kappa_{\mathrm{L}}$ below 1 Wm$^{-1}$K$^{-1}$, and most have values below 2 Wm$^{-1}$K$^{-1}$. Among the evaluated high-$\kappa_{\mathrm{L}}$ candidates, 44\% exhibit $\kappa_{\mathrm{L}}$ above 100 Wm$^{-1}$K$^{-1}$. These results demonstrate that the combination of MSD and $\omega_{\min}^{\mathrm{opt}}$ provides an effective, physically interpretable route to the large-scale discovery of materials with ultralow and ultrahigh $\kappa_{\mathrm{L}}$.

\section{Methods}
\subsection{MatterSim Potential Refinement}
Because the descriptors used in this study, MSD and $\omega_{\min}^{\mathrm{opt}}$, are primarily determined by harmonic interatomic force constants, an accurate description of harmonic lattice dynamics is essential. Accordingly, we fine-tuned the machine-learning interatomic potential MatterSim~\cite{yang2024mattersimdeeplearningatomistic} using second-order interatomic force constants (IFC2) obtained from PhononDB~\cite{TOGO20151}, which provides DFT-calculated IFC2 for approximately 10,000 inorganic crystalline materials. After data cleaning, 10,031 structures with usable IFC2 data were retained for this study. The training workflow is illustrated in Fig.~\ref{fig:filter_step}, and full implementation details are provided in Section~\ref{Training_details} of the Supplementary Information.

\subsection{Screening Strategy for Materials}
The validated model was applied to a large set of candidate materials from the Open Quantum Materials Database (OQMD)~\cite{OQMD}. Materials were considered for further analysis only if they had a band gap greater than 0.1 eV, a formation energy convex-hull distance below 0.1 eV/atom, and contained no hydrogen, noble-gas, or radioactive elements. For each compound, IFC2 were calculated using the fine-tuned MatterSim potential (hereafter referred to as Fit-MatterSim). The resulting force constants were used as input to \textsc{PHONOPY}~\cite{PhysRevB.77.144112,TOGO20151} to calculate the phonon frequencies and MSDs of the constituent atoms. Compounds with significant imaginary phonon modes were excluded to ensure dynamical stability and reliable MSD values. For each retained compound, $\omega_{\min}^{\mathrm{opt}}$ was extracted, and the overall MSD was defined as the arithmetic mean of the atomic MSDs in the compound~\cite{Rodriguez2026}. This average provides a measure of the overall atomic vibrational amplitude and facilitates comparisons among compounds with different numbers of atoms. Unless otherwise specified, MSD refers hereafter to the average over all atoms in a compound. Candidate materials with extreme thermal transport properties were identified based on the calculated MSD and $\omega_{\min}^{\mathrm{opt}}$. A detailed analysis is provided in Section~\ref{screening}.

\subsection{COMPUTATION  DETAILS}\label{computing details}
For compounds selected based on their MSD and $\omega_{\min}^{\mathrm{opt}}$, DFT calculations were performed to evaluate their $\kappa_{\mathrm{L}}$. All DFT calculations were carried out using the Vienna \textit{Ab initio} Simulation Package (\textsc{VASP})~\cite{PhysRevB.54.11169,vasp2}. The interactions between valence and core electrons were described using the projector augmented-wave method~\cite{PhysRevB.50.17953,PAW2}. The PBEsol exchange--correlation functional~\cite{PhysRevLett.100.136406} was employed, as it has been shown to perform well for $\kappa_{\mathrm{L}}$ calculations~\cite{PhysRevB.110.035205}. A plane-wave basis set with a kinetic-energy cutoff of 520~eV and a Monkhorst--Pack $k$-point mesh were used. All crystal structures were fully relaxed until the residual force on each atom was below 0.01~eV/\AA, with a total-energy convergence criterion of $10^{-5}$~eV. The IFC2 were calculated using the finite-displacement method implemented in \textsc{Phonopy} code~\cite{PhysRevB.77.144112,TOGO20151}. To improve the accuracy of the force calculations, the reciprocal-space projection scheme was employed, together with a stringent electronic convergence threshold of $10^{-8}$~eV. Higher-order force constants were obtained using the compressive sensing lattice dynamics (CSLD) method~\cite{PhysRevLett.113.185501} implemented in the \textsc{ALAMODE} package~\cite{Tadano_2014}. The IFC2 were renormalized using the self-consistent phonon (SCPH) method~\cite{PhysRevLett.106.165501,PhysRevLett.120.105901}. Lattice thermal transport properties were calculated by iteratively solving the Peierls--Boltzmann transport equation using the \textsc{FourPhonon} package~\cite{HAN2022108179}. Wigner coherent conductivity was evaluated within the framework of the unified theory of thermal transport~\cite{Simoncelli2019}. The computational workflow is illustrated in Fig.~\ref{fig:filter_step}.

\begin{figure*}[tph!]
	\includegraphics[clip,width=1.0\linewidth]{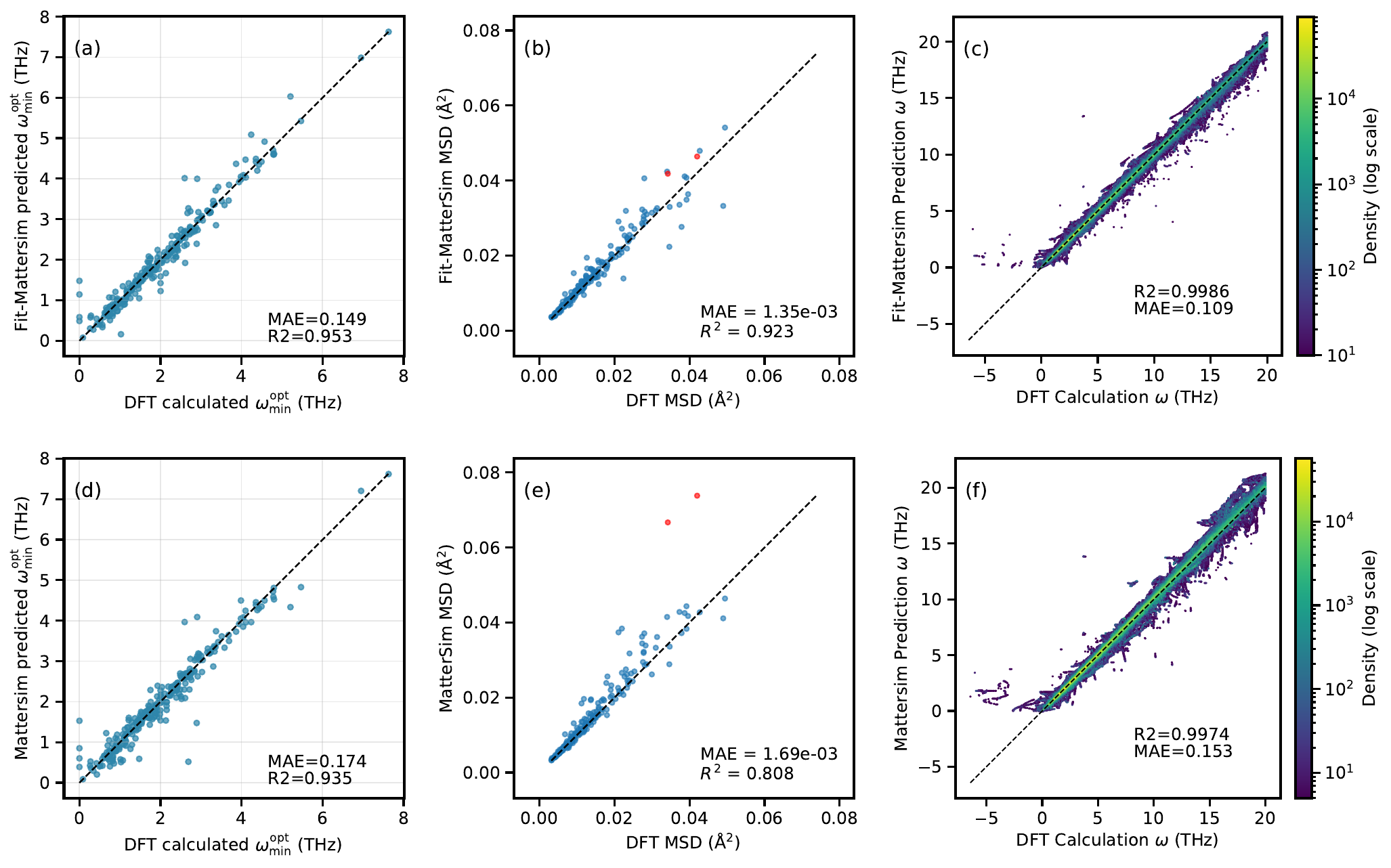}
	\caption{Comparison of properties predicted by the fine‑tuned Fit‑MatterSim (top row) and original MatterSim (bottom row) against DFT references in the test set. (a,d) Comparison of minimum optical phonon frequencies. (b,e) Comparison of mean‑squared displacement (MSD) values. (c,f) Overall phonon‑frequency distribution, where colour intensity represents the density of data points.}
	\label{fig:lwof_msd_mae}
\end{figure*}

\section{RESULTS AND DISCUSSION}
\subsection{Physical Descriptors of Lattice Thermal Conductivity}
Despite substantial progress in understanding phonon-transport mechanisms and developing high-throughput screening strategies, the efficient identification of materials with extreme $\kappa_{\mathrm{L}}$ remains challenging. Accurate screening typically relies either on direct calculations of $\kappa_{\mathrm{L}}$, which are computationally demanding, or on data-driven models that require sufficiently large and reliable labeled datasets. Moreover, many previously proposed descriptors are applicable only to specific structural families or chemical spaces. Consequently, simple, physically interpretable, and broadly applicable descriptors capable of distinguishing materials with exceptionally low or high $\kappa_{\mathrm{L}}$ remain highly desirable. The MSD quantifies the mean-squared amplitude of atomic vibrations about the equilibrium positions and provides an indication of local lattice softness. At a given temperature, larger MSD values generally indicate weaker atomic restoring forces and softer vibrational modes~\cite{10.1002/advs.202417292,z9b4-y2fl}. Such weakly constrained atomic vibrations are often associated with phonon branches having lower $\nu_{\mathrm{g}}$, pronounced lattice anharmonicity, and enhanced phonon scattering, which can collectively suppress $\kappa_{\mathrm{L}}$. Conversely, smaller MSD values can indicate stronger local restoring forces and a more rigid lattice, which may support higher group velocities and promote efficient heat transport~\cite{10.1063/5.0201778,SALES1999528,SLACK19791}.

Another physically relevant quantity is the minimum optical phonon frequency, $\omega_{\min}^{\mathrm{opt}}$~\cite{z9b4-y2fl}, which characterizes the lowest energy scale of optical vibrations. Such low-frequency modes can arise from weak interatomic interactions or large atomic masses~\cite{PhysRevB.110.035205,PhysRevB.108.024306}. A low $\omega_{\min}^{\mathrm{opt}}$ can bring optical modes into the acoustic-frequency range, where a substantial portion of lattice heat transport occurs, and may enlarge the scattering phase space by opening additional acoustic--optical phonon-scattering channels that satisfy energy and crystal-momentum conservation~\cite{D2TC03356A,PhysRevB.109.115202}. Coupling between these low-frequency optical modes and acoustic modes may also reduce the group velocities of acoustic phonons through avoided crossings~\cite{10.1002/inf2.12217}. These effects can collectively suppress $\kappa_{\mathrm{L}}$ through reduced $\nu_{\mathrm{g}}$ and enhanced phonon--phonon scattering rates, $1/\tau$. In contrast, a relatively high $\omega_{\min}^{\mathrm{opt}}$ increases the spectral separation between the acoustic and optical modes, restricts certain acoustic--optical scattering channels, and may therefore promote higher $\kappa_{\mathrm{L}}$~\cite{PhysRevB.109.115202,10.1002/adfm.202108532}. Thus, the combination of MSD and $\omega_{\min}^{\mathrm{opt}}$ may provide complementary dynamical descriptors for rapidly identifying materials at both the low-$\kappa_{\mathrm{L}}$ and high-$\kappa_{\mathrm{L}}$ extremes.

Although MSD and $\omega_{\min}^{\mathrm{opt}}$ provide physically interpretable descriptors of $\kappa_{\mathrm{L}}$, calculating these quantities using DFT remains computationally expensive, particularly when screening thousands to tens of thousands of candidate materials. In recent years, the development of machine-learning interatomic potentials has enabled accurate predictions of atomic forces and phonon properties at a fraction of the computational cost of DFT. The use of machine-learning potentials enables the efficient calculation of MSDs and phonon frequencies across large-scale materials datasets, thereby facilitating the rapid screening of promising candidates with potentially ultralow or ultrahigh $\kappa_{\mathrm{L}}$.

\subsection{Benchmarking the Refined Potential}
Conventional machine-learning interatomic potentials are typically trained using total energies and atomic forces. Although these training targets enable accurate structural relaxation and molecular-dynamics simulations, high accuracy in energy and force predictions does not necessarily ensure an equally accurate description of harmonic lattice dynamics. Total energies characterize the potential-energy surface, whereas forces describe its first-order response to atomic displacements. Harmonic phonon properties, in contrast, are governed directly by the second-order response of the potential-energy surface.

The IFC2 quantify the change in the force acting on one atom in response to the displacement of another. They determine the dynamical matrix and, consequently, the phonon dispersion, dynamical stability, and harmonic vibrational properties, including the MSD. Thus, IFC2 provide more direct information about lattice dynamics than total energies and atomic forces alone. To improve the description of harmonic lattice dynamics, MatterSim was fine-tuned using the large-scale IFC2 dataset from PhononDB. The resulting model, denoted Fit-MatterSim, was optimized primarily against IFC2, with force terms retained as auxiliary constraints to help preserve the accuracy of force predictions. In this section, the performance of Fit-MatterSim is benchmarked against that of the original MatterSim model by directly comparing the predicted IFC2 and the resulting phonon frequencies. A total of 10,031 compound--force-constant pairs were collected from the PhononDB dataset. Their elemental and crystal-system distributions are shown in Fig.~\textcolor{darkteal}{S1}. \ignore{arg1\ref{fig_SI:train_ele_sg_dist}}

Before fine-tuning, 1,004 compounds were randomly selected as an independent test set for evaluating the final performance of Fit-MatterSim. The remaining 9,027 structures were randomly divided into training and validation sets in an 85:15 ratio. To further assess model performance, we compared the $\omega_{\min}^{\mathrm{opt}}$, MSD, and phonon-frequency distributions predicted by the original MatterSim and Fit-MatterSim with the corresponding DFT results for the test set, as shown in Fig.~\ref{fig:lwof_msd_mae}. Compounds exhibiting significant imaginary phonon modes in the predictions of either MatterSim or Fit-MatterSim were excluded according to the criteria in Table~\textcolor{darkteal}{S1}. After applying these criteria, 396 compounds remained for comparison. Imaginary modes may arise from intrinsic dynamical instability or from inaccuracies in the model's description of the potential-energy surface.

Figure~\ref{fig:lwof_msd_mae}(a) and (d) compare the $\omega_{\min}^{\mathrm{opt}}$ values predicted by the Fit-MatterSim and original MatterSim, respectively, with the corresponding DFT results. Fine-tuning reduces the MAE from 0.174 to 0.149~THz and increases the $R^2$ from 0.935 to 0.953. This modest improvement is attributable primarily to the high accuracy of the original MatterSim predictions for $\omega_{\min}^{\mathrm{opt}}$, which leaves limited scope for further improvement. Fig.~\ref{fig:lwof_msd_mae}(b) and (e) show the corresponding comparisons for MSD. Fine-tuning substantially improves predictive accuracy, increasing the $R^2$ from 0.808 to 0.923. This improvement is accompanied by a marked reduction in prediction errors for two prominent outliers with relatively large MSD values. More generally, structures with larger MSD values tend to have larger prediction errors, potentially reflecting greater deviations from the harmonic approximation. Fit-MatterSim reduces these deviations for high-MSD structures, as discussed in detail in Section~\ref{section_descriptor_select}. Figure~\ref{fig:lwof_msd_mae}(c) and (f) compare the predicted phonon frequencies across the full phonon spectrum. Here, the increase in $R^2$ is relatively small, likely because this aggregate metric is dominated by the large number of frequencies that the original MatterSim already reproduces accurately. Consequently, correcting a relatively small fraction of predictions with large errors has only a limited effect on the overall $R^2$. By contrast, fine-tuning substantially reduces the MAE, from 0.153 to 0.109~THz. This reduction, together with the smaller number of points lying far from the diagonal, indicates that Fit-MatterSim reduces phonon-frequency prediction errors, particularly for frequencies that are poorly described by the original MatterSim.

Overall, fine-tuning yields relatively modest improvements in the global statistical metrics, primarily because the original MatterSim already predicts most of the lattice-dynamical quantities considered here with high accuracy. Nevertheless, Fit-MatterSim partially corrects predictions for structures with relatively large errors in the original model, thereby improving the robustness of its description of lattice-dynamical properties. In particular, fine-tuning improves the reliability of the descriptors used in our screening, as discussed in Section~\ref{section_descriptor_select}. Accordingly, Fit-MatterSim was used in the subsequent screening calculations to evaluate descriptors for identifying materials with low and high $\kappa_{\mathrm{L}}$.

\subsection{Descriptor Selection for Lattice Thermal Conductivity}\label{section_descriptor_select}
\begin{figure*}[tph!]
	\includegraphics[clip,width=1.05\linewidth]{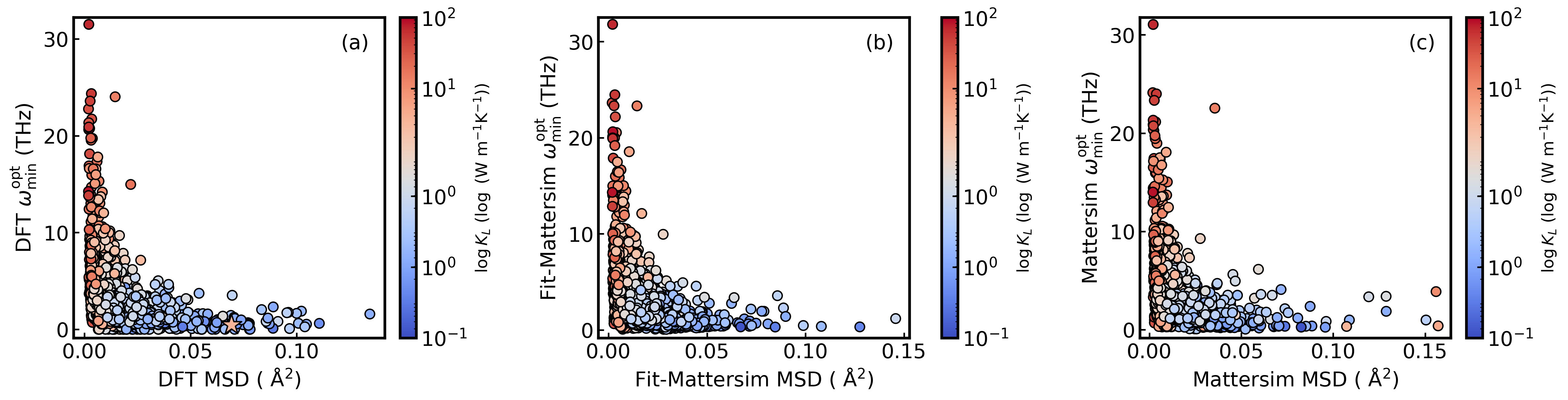}
	\caption{Correlations between room-temperature $\kappa_{\mathrm{L}}$ and MSD and $\omega_\text{min}^{\mathrm{opt}}$. All the $\kappa_{\mathrm{L}}$ value are from Phonix~\cite{WOS:001739467100001} database, and the compounds with nonphysical $\kappa_{\mathrm{L}}$ value (i.e., $<$ 10$^{-3}$ or > 3000 Wm$^{-1}$K$^{-1}$) are excluded. (a) DFT calculated MSD and $\omega_{\mathrm{min}}^{\mathrm{opt}}$; (b) Fit‑MatterSim predicted MSD and $\omega_{\mathrm{min}}^{\mathrm{opt}}$; (c) original MatterSim predicted MSD and $\omega_{\mathrm{min}}^{\mathrm{opt}}$. Structures with substantial imaginary phonon modes are excluded in panels (b) and (c).}
	\label{val_msd_freq}
\end{figure*}

As discussed above, compounds with large MSD and low $\omega_{\min}^{\mathrm{opt}}$ are generally expected to exhibit low $\kappa_{\mathrm{L}}$, whereas those with small MSD and high $\omega_{\min}^{\mathrm{opt}}$ are more likely to exhibit high $\kappa_{\mathrm{L}}$. To further examine this trend, we collected crystal structures with both second-order interatomic force constants and $\kappa_{\mathrm{L}}$ data from the Phonix~\cite{WOS:001739467100001} database and calculated their MSD and $\omega_{\min}^{\mathrm{opt}}$. As shown in Fig.~\ref{val_msd_freq}(a), large MSD and low $\omega_{\min}^{\mathrm{opt}}$ are generally associated with low $\kappa_{\mathrm{L}}$, whereas the opposite combination is associated with high $\kappa_{\mathrm{L}}$, further supporting the preceding analysis. Notably, SiI$_4$ ($Pa\overline{3}$, No. 205), marked by a star in Fig.~\ref{val_msd_freq}(a), deviates substantially from this trend. SiI$_4$ has a relatively large MSD and a low $\omega_{\min}^{\mathrm{opt}}$, yet the Phonix~\cite{WOS:001739467100001} database reports a high $\kappa_{\mathrm{L}}$ of 32.16 Wm$^{-1}$K$^{-1}$ at 300 K. We therefore recalculated its $\kappa_{\mathrm{L}}$ and obtained a value of only 0.28 Wm$^{-1}$K$^{-1}$ at 300 K, substantially below the reported value. With this recalculated $\kappa_{\mathrm{L}}$, SiI$_4$ conforms to the trend. This result provides further support for the relationship between $\kappa_{\mathrm{L}}$, MSD, and $\omega_{\min}^{\mathrm{opt}}$. Although exceptions may occur because thermal transport depends on the full phonon spectrum and the associated scattering processes, this complementary set of descriptors provides a practical guideline for high-throughput screening of materials with targeted thermal transport properties.

Although the trends described above are grounded in physical principles, Fit-MatterSim, as a machine-learning interatomic potential, may introduce uncertainties in the calculated MSD and $\omega_{\min}^{\mathrm{opt}}$ relative to first-principles results. We therefore used compounds retrieved from the Phonix~\cite{WOS:001739467100001} database to assess whether Fit-MatterSim can reproduce the expected correlations of these descriptors with $\kappa_{\mathrm{L}}$. Because both the Phonix~\cite{WOS:001739467100001} dataset and the PhononDB~~\cite{TOGO20151} dataset used to fine-tune MatterSim contain structures originating from the Materials Project~\cite{Horton2025}, we removed compounds shared by the two datasets to avoid data overlap. We also excluded compounds for which Fit-MatterSim predicted significant imaginary phonon modes. These modes were identified using combined criteria based on the number of imaginary-frequency points and the magnitudes of the imaginary frequencies, as detailed in Table~\textcolor{darkteal}{S1}\ignore{\ref{tab:phonon_filter_criteria}}. Such modes may arise either from intrinsic dynamical instability or from limitations of the machine-learning potential in describing the interatomic force constants. After removing overlapping compounds and those with significant imaginary phonon modes, 1,550 compounds remained for quantitative comparison. We compared the MSD values at 300 K and $\omega_{\min}^{\mathrm{opt}}$ predicted by Fit-MatterSim for these compounds with the corresponding DFT-derived values reported in the Phonix~\cite{WOS:001739467100001} database, as shown in Fig.~\textcolor{darkteal}{S3}.

As shown in Fig.~\textcolor{darkteal}{S3}\ignore{(\ref{fig_SI:fit_dft_msd_freq_comparison})}, Fit-MatterSim predicts $\omega_{\min}^{\mathrm{opt}}$ with high accuracy, achieving an $R^2$ of 0.978 and a low MAE of 0.2259 THz. Its predictive accuracy for MSD is lower, with an $R^2$ of 0.7526. We also compared the MSD at 300 K and $\omega_{\min}^{\mathrm{opt}}$ predicted by the original MatterSim model with the DFT results. As shown in Fig.~\textcolor{darkteal}{S3}\ignore{\ref{fig_SI:fit_dft_msd_freq_comparison}} and Fig.~\textcolor{darkteal}{S4}\ignore{\ref{fig_SI:mat_dft_msd_freq_comparison}}, incorporating second-order force-constant supervision improves the predictions of both $\omega_{\min}^{\mathrm{opt}}$ and MSD. For MSD, the $R^2$ increases from 0.5416 to 0.7526, while the MAE decreases from 0.0024\ $\mathrm{\text{\AA}^2}$ to 0.0019\ $\mathrm{\text{\AA}^2}$. The improvement in $\omega_{\min}^{\mathrm{opt}}$ is marginal because the original MatterSim model already predicts this quantity with high accuracy. These results indicate that including second-order force-constant information enhances the model's ability to capture lattice dynamics and vibrational behavior.

As shown in Fig.~\textcolor{darkteal}{S3}\ignore{\ref{fig_SI:fit_dft_msd_freq_comparison}}(a) and Fig.~\textcolor{darkteal}{S4}\ignore{\ref{fig_SI:mat_dft_msd_freq_comparison}}(a), the MSD prediction errors of both Fit-MatterSim and the original MatterSim model generally increase with the DFT-calculated MSD. This behavior is physically reasonable: a large MSD generally indicates weak restoring forces and a small effective force constant around the equilibrium configuration~\cite{10.1063/1.5135584}. For a given atomic displacement, such a soft potential-energy surface produces only a small restoring force. Consequently, an absolute error of a given magnitude in the predicted force response can correspond to a larger relative error in the effective force constant when the true restoring force is small. Under the harmonic approximation and in the classical high-temperature limit, the MSD scales approximately inversely with the effective restoring force constant~\cite{housley1966analysis}. Thus, the larger relative error in the effective force constant of a soft structure can translate directly into a larger error in the predicted MSD.

For the original MatterSim model, both overestimation and underestimation occur in the high-MSD region. MatterSim was trained on energies, forces, and stresses for a diverse set of equilibrium and off-equilibrium configurations generated through active learning~\cite{yang2024mattersimdeeplearningatomistic}. Although this training constrains the potential-energy surface across a broad configurational space, it does not directly supervise the local harmonic curvature. Residual errors in the curvature can therefore occur in either direction, consistent with the observed overestimation and underestimation of MSD. Underestimating the effective restoring forces permits excessively large thermal displacements and leads to an overestimated MSD, whereas overestimating them constrains atomic motion too strongly and results in an underestimated MSD~\cite{10.1002/aidi.202500075}. Following fine-tuning with IFC2 data, the overall MSD predictions improve substantially. Nevertheless, relatively large errors remain in the high-MSD region, where the predicted MSD values are predominantly below the DFT references. This residual underestimation may arise because the weak restoring-force components that govern large MSD values in soft structures make up only a small fraction of the force-constant data used for fine-tuning. Since most of the training data represent more typical restoring-force magnitudes, these exceptionally weak responses lie in the tail of the training distribution. During global optimization, predictions for such extreme values may therefore shift toward the more frequently represented force-constant scale, resulting in a slight overestimation of the effective restoring forces in the softest structures. Because the MSD is highly sensitive to weak restoring forces, this overestimation can lead to the systematic underestimation of MSD observed in the high-MSD region.

Importantly, this systematic underestimation is concentrated mainly in the high-MSD region and is therefore not expected to substantially compromise subsequent screening. High-$\kappa_{\mathrm{L}}$ candidates are selected using a small-MSD criterion, within which Fit-MatterSim agrees much better with the DFT results. For low-$\kappa_{\mathrm{L}}$ screening, the observed underestimation represents a conservative bias: a material that meets the large-MSD cutoff according to the Fit-MatterSim prediction is likely to have a comparable or even larger DFT-calculated MSD, further supporting its identification as a low-$\kappa_{\mathrm{L}}$ candidate. This bias may cause some materials close to the cutoff to be overlooked, thereby reducing screening recall, but it is unlikely to substantially reduce the reliability of the retained candidates. In contrast, the coexistence of MSD overestimation and underestimation in the original MatterSim predictions increases the risk of both false-positive and false-negative selections, making these predictions less suitable for defining a reliable screening cutoff.

To examine whether the descriptors derived from Fit-MatterSim retain the expected relationships with $\kappa_{\mathrm{L}}$, we used the same set of compounds from the Phonix~\cite{WOS:001739467100001} database as in the analysis presented in Fig.~\ref{val_msd_freq}(a). We calculated their MSD and $\omega_{\min}^{\mathrm{opt}}$ using Fit-MatterSim. After excluding compounds with significant imaginary phonon modes predicted by Fit-MatterSim, according to the criteria listed in Table~\textcolor{darkteal}{S1}\ignore{\ref{tab:phonon_filter_criteria}} of the Supporting Information, 5,742 materials remained. Figure~\ref{val_msd_freq}(b) shows the relationships between the Fit-MatterSim-predicted MSD at 300 K and $\omega_{\min}^{\mathrm{opt}}$, and the $\kappa_{\mathrm{L}}$ at 300 K reported in the Phonix~\cite{WOS:001739467100001} database. The overall trends are consistent with the physical expectations discussed above, supporting the use of MSD and $\omega_{\min}^{\mathrm{opt}}$ calculated using Fit-MatterSim as efficient descriptors for screening materials with targeted lattice thermal conductivities.


Figure~\ref{val_msd_freq}(c) shows the relationships between the MSD and $\omega_{\min}^{\mathrm{opt}}$ predicted by the original MatterSim model and the $\kappa_{\mathrm{L}}$ values reported in the Phonix~\cite{WOS:001739467100001} database, using the same screening criteria as in Fig.~\ref{val_msd_freq}(b). As shown in Fig.~\ref{val_msd_freq}(c), the original MatterSim model predicts a noticeable number of materials with both large MSD and high $\kappa_{\mathrm{L}}$. Although such combinations are not impossible, given that $\kappa_{\mathrm{L}}$ depends on multiple factors, their frequent occurrence is inconsistent with the expected overall relationship between MSD and $\kappa_{\mathrm{L}}$. This discrepancy is even more evident when these predictions are compared with the DFT-calculated results in Fig.~\ref{val_msd_freq}(a). In the DFT results, materials identified by the original MatterSim model as having both large MSD and high $\kappa_{\mathrm{L}}$ do not show the same anomalous behavior; such combinations are absent from the relationships between $\kappa_{\mathrm{L}}$, MSD, and $\omega_{\min}^{\mathrm{opt}}$.

In contrast, the Fit-MatterSim results contain substantially fewer such anomalies and are more consistent with the DFT results. These comparisons indicate that incorporating second-order force-constant information improves the model's description of lattice vibrational properties and substantially reduces the anomalous predictions observed with the original MatterSim model. Based on these observations, we use MSD and $\omega_{\min}^{\mathrm{opt}}$ calculated by Fit-MatterSim as descriptors to screen materials for extreme $\kappa_{\mathrm{L}}$, enabling efficient exploration of large materials spaces.

\begin{figure}[tph!]
	\includegraphics[clip,width=1.0\linewidth]{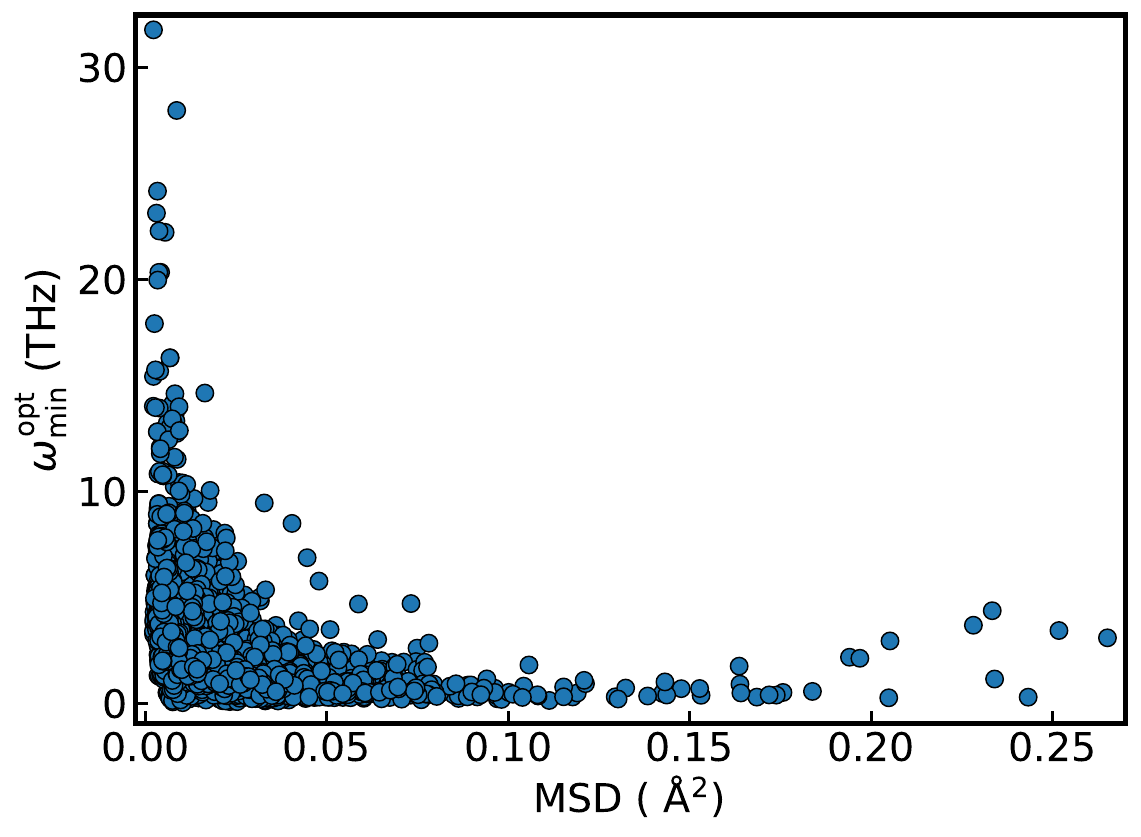}
	\caption{MSD versus the  $\omega_{\min}^{\mathrm{opt}}$ for the 8,974 filtered structures.}
	\label{fig:scatter_msd_freq_mean}
\end{figure}

\begin{figure}[tph!]
	\includegraphics[clip,width=1.0\linewidth]{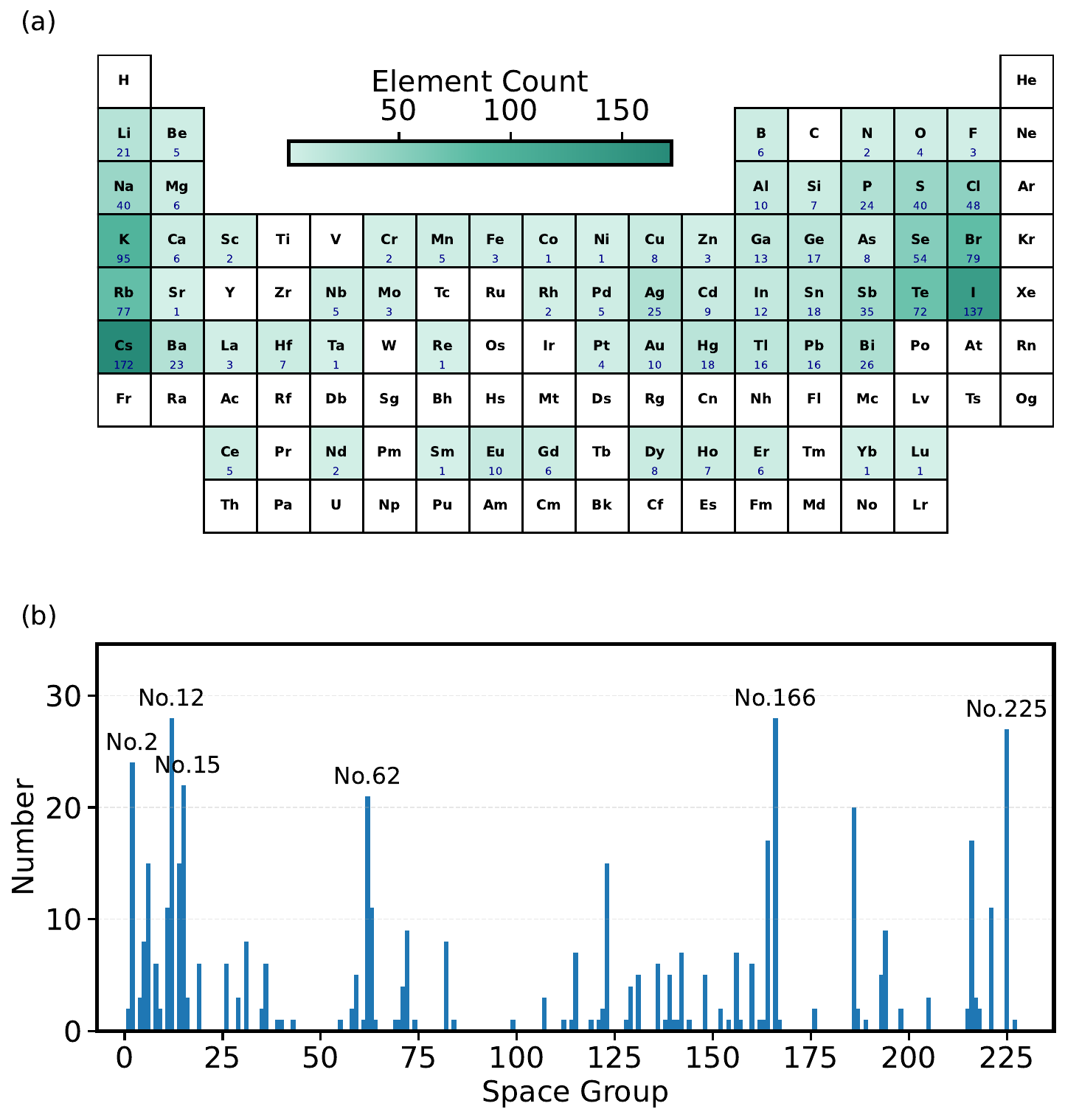}
	\caption{Statistical distribution of the screened low lattice thermal conductivity candidates, with a total of 463 structures. (a) Elemental abundance of the screened candidates, where each element is counted only once per structure regardless of its atomic occupancy within a single compound. (b) Distribution of the dominant space groups for the screened low‑thermal‑conductivity structures. }
	\label{fig:ele_sg_dist_lower}
\end{figure}

\begin{figure}[tph!]
	\includegraphics[clip,width=1\linewidth]{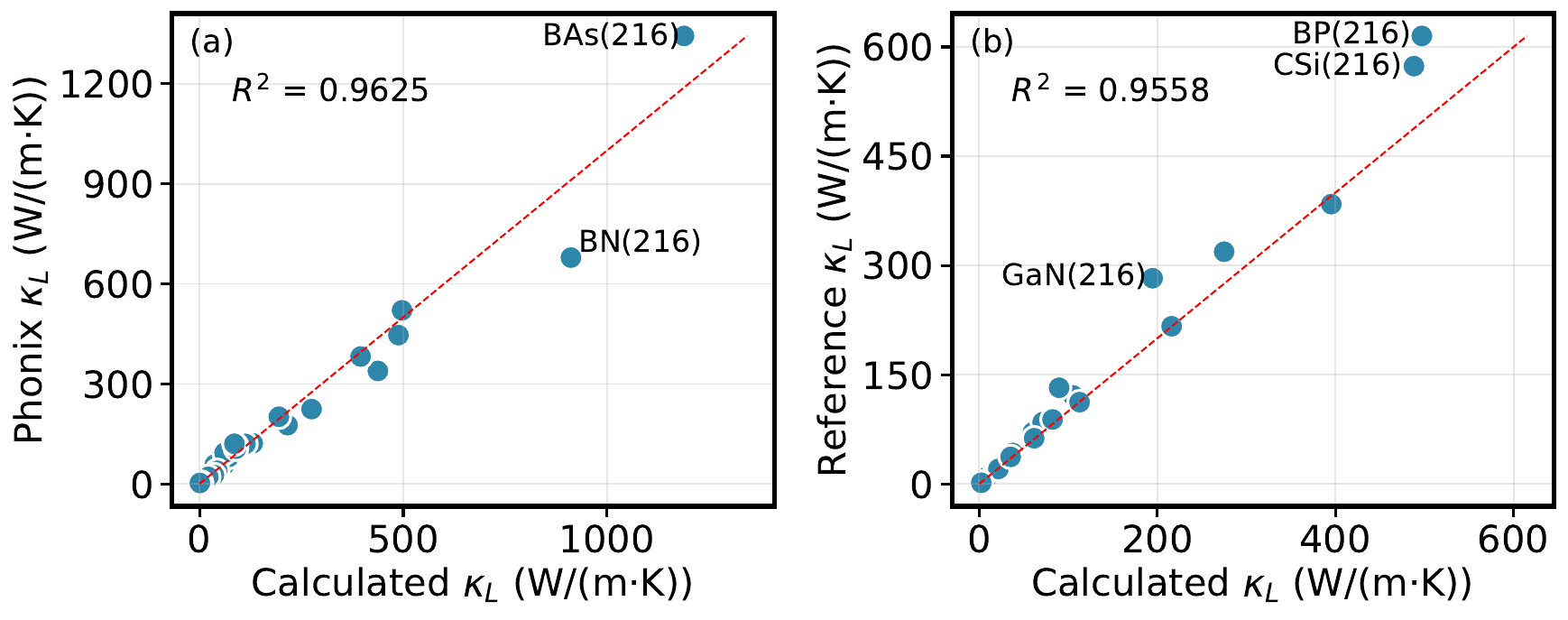}
	\caption{Benchmark of calculated lattice thermal conductivity against literature and database results. (a) Correlation plot comparing our calculated lattice thermal conductivities with values from the Phonix database. (b) Comparison of our computed lattice thermal conductivities with the computational results reported by Li et al.}
	\label{fig:ltc_compare}
\end{figure}

\begin{figure}[tph!]
	\includegraphics[clip,width=1.0\linewidth]{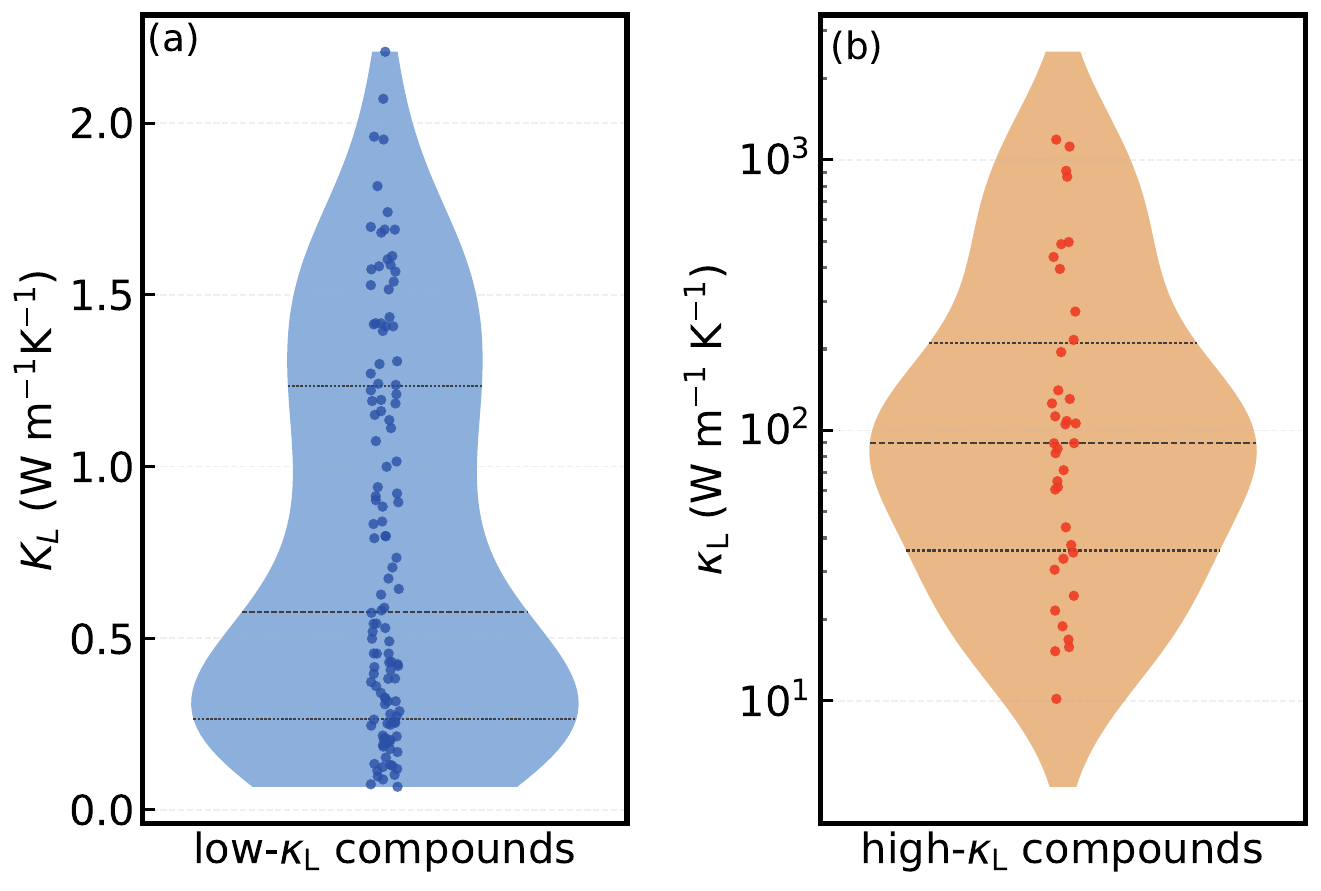}
	\caption{Violin plots of the lattice thermal conductivity distribution for DFT‑calculated structures. (a) Distribution of lattice thermal conductivity for low $\kappa_{L}$ materials obtained from DFT calculations. (b) Distribution of lattice thermal conductivity for high $\kappa_{L}$ materials obtained from DFT calculations.}
	\label{fig:ltc_distribution}
\end{figure}

\subsection{Fit-MatterSim Accelerated Screening of Materials with Extreme-$\kappa_{\mathrm{L}}$}\label{screening}
In order to explore a sufficiently broad chemical space, we screened compounds from the Open Quantum Materials Database (OQMD)~\cite{OQMD}, which contains more than one million inorganic crystalline compounds with calculated thermodynamic stability and band-gap information. Because our focus is on synthesizable semiconductors, we applied a three-step filtering procedure. First, we retained compounds that were thermodynamically stable (on the formation-energy convex hull) or metastable (with a convex-hull distance below 0.1 eV/atom). Second, we selected compounds with a PBE-calculated band gap greater than 0.1~eV to exclude metals. Third, we excluded compounds containing hydrogen, radioactive elements, or noble gases. Duplicate entries were then removed based on chemical composition and space group, yielding a dataset of 114,580 compounds for subsequent screening. For these compounds, second-order interatomic force constants were calculated using Phonopy~\cite{phonopy-phono3py-JPCM,phonopy-phono3py-JPSJ}, with forces calculated using Fit-MatterSim. The corresponding $\omega_{\min}^{\mathrm{opt}}$ and MSD values at 300 K were then obtained using Phonopy~\cite{phonopy-phono3py-JPCM,phonopy-phono3py-JPSJ}. As described above, compounds exhibiting significant imaginary phonon modes were excluded according to the criteria summarized in Table~\textcolor{darkteal}{S1}\ignore{\ref{tab:phonon_filter_criteria}}. After this filtering step, 8,974 compounds remained. To identify potential materials with extremely high or low $\kappa_{\mathrm{L}}$, we used the previously established relationships between $\kappa_{\mathrm{L}}$, MSD, and $\omega_{\min}^{\mathrm{opt}}$. As shown in Fig.~\ref{fig:scatter_msd_freq_mean}, the compound distribution is similar to that in Fig.~\ref{val_msd_freq}: most compounds lie in the region of low MSD and low $\omega_{\min}^{\mathrm{opt}}$, while compounds with high $\omega_{\min}^{\mathrm{opt}}$ generally exhibit low MSD, and vice versa.

As shown in Fig.~\ref{val_msd_freq}, materials with an MSD at 300~K greater than 0.04~\AA$^2$ and $\omega_{\min}^{\mathrm{opt}}$ below 5 THz generally exhibit low $\kappa_{\mathrm{L}}$, with most having $\kappa_{\mathrm{L}} \leq 1.0$ Wm$^{-1}$K$^{-1}$. Conversely, materials with an MSD below 0.02~\AA$^2$ and $\omega_{\min}^{\mathrm{opt}} \geq 10$ THz generally exhibit high $\kappa_{\mathrm{L}}$, with most having $\kappa_{\mathrm{L}} > 100$ Wm$^{-1}$K$^{-1}$. Accordingly, we identified materials satisfying MSD > 0.04~\AA$^2$ and $\omega_{\min}^{\mathrm{opt}}$ < 5 THz as potential low-$\kappa_{\mathrm{L}}$ candidates, and those satisfying MSD < 0.02~\AA$^2$ and $\omega_{\min}^{\mathrm{opt}}$ > 10 THz as potential high-$\kappa_{\mathrm{L}}$ candidates. Applying these criteria yielded 463 low-$\kappa_{\mathrm{L}}$ candidates and 49 high-$\kappa_{\mathrm{L}}$ candidates.

Figure~\ref{fig:ele_sg_dist_lower} shows the elemental and space-group distributions of the low-$\kappa_{\mathrm{L}}$ candidates. Cs and I are the two most frequently occurring elements, appearing in 172 and 137 candidate compounds, respectively, despite not being particularly prevalent in the initial dataset, as shown in Fig.~\textcolor{darkteal}{S2}\ignore{\ref{fig_SI:ele_sg_dist_all}}(a). This enrichment may reflect the combined effects of large atomic masses and soft local bonding environments. Cs has a low electronegativity and readily loses its outermost 6$s$ electron to form a Cs$^{+}$ cation, which commonly serves as a charge-compensating cation in multinary compounds. Cs has a larger atomic mass than Rb; compared with Ba$^{2+}$, which has a similar atomic mass and electronegativity, Cs$^{+}$ has a smaller formal charge, and thus weaker Coulomb interactions, as well as a larger ionic radius at the same coordination number. In comparable ionic coordination environments, its low formal charge and relatively long coordination distances may favor weak local bonding and large MSD. These weak restoring forces, together with its large atomic mass, also favor low-frequency optical phonon modes~\cite{10.1002/adfm.202304607,cao2602.06501}. Iodine typically occurs as a monovalent anion in metal iodides. Compared with Br, I has a larger atomic mass and ionic radius but lower electronegativity. Consequently, metal--iodide bonds are longer than metal--bromide bonds and may have weaker restoring forces. The diffuse 5$p$ electron cloud of I is also highly polarizable, which may further contribute to lattice softening~\cite{10.1021/jp068257s,10.1021/jacs.7b06327}. The resulting weak restoring forces favor large MSD and, together with I's large atomic mass, low-frequency optical phonon modes. Te also has a large atomic mass and a polarizable valence electron cloud, but commonly adopts a formal oxidation state of $-2$ in tellurides. Within an ionic model with comparable cations, interionic distances, and coordination environments, the smaller formal charge magnitude of I$^{-}$ corresponds to weaker electrostatic interactions than that of Te$^{2-}$. In addition, I is more electronegative than Te, favoring greater ionic character when bonded to the same electropositive metal. Actual bonding characteristics, however, also depend on orbital hybridization, electronic-state occupation, and the local coordination environment~\cite{lee2014resonant}. Together, these considerations provide a qualitative explanation for the preferential selection of Cs- and I-containing compounds by screening criteria based on large MSD and low $\omega_{\min}^{\mathrm{opt}}$.

In contrast, Fig.~\textcolor{darkteal}{S5}\ignore{\ref{fig_SI:ele_sg_dist_high}}(a) shows the elemental distribution of the candidates screened for high $\kappa_{\mathrm{L}}$. Light elements such as Li, Be, N, and O are prominent, whereas Cs, I, Rb, and Br, which are abundant among the low-$\kappa_{\mathrm{L}}$ candidates, are entirely absent. This contrast further supports the effectiveness of our screening strategy for identifying low- and high-$\kappa_{\mathrm{L}}$ candidates.

As shown in Fig. \ref{fig:ele_sg_dist_lower}(b), the screened low-$\kappa_{\mathrm{L}}$ candidates span a broad range of space groups. The most common are $P\overline 1$ (No. 2), $C2/m$ (No. 12), $C2/c$ (No. 15), $Pnma$ (No. 62), $R\overline{3}m$ (No. 166), and $Fm\overline{3}m$ (No. 225). Several of these space groups, particularly $Pnma$, $R\overline{3}m$, and $C2/m$, are also well represented in the screening dataset, and their relative rankings change only modestly after screening. This broad distribution is consistent with the multiple factors that govern $\kappa_{\mathrm{L}}$. Even highly symmetric crystals can exhibit low $\kappa_{\mathrm{L}}$ because of heavy constituent atoms, weak chemical bonding, or strong phonon scattering~\cite{cao2602.06501,10.1002/adfm.202108532,PhysRevB.92.020301}.

By contrast, the high-$\kappa_{\mathrm{L}}$ candidates are concentrated in relatively high-symmetry space groups. $F\overline{4}3m$ (No. 216) is the most prevalent, followed by $R\overline{3}m$ and $Fm\overline{3}m$, as shown in Fig.~\textcolor{darkteal}{S5}\ignore{\ref{fig_SI:ele_sg_dist_high}}. This narrower distribution suggests a more selective structural preference within the screened dataset. Lower-symmetry structures may exhibit greater structural complexity or stronger anharmonicity, either of which can suppress $\kappa_{\mathrm{L}}$ by reducing phonon group velocities or enhancing phonon scattering~\cite{Chen2025}. High symmetry is not, however, a prerequisite for high $\kappa_{\mathrm{L}}$. As shown in Fig.~\textcolor{darkteal}{S5}\ignore{\ref{fig_SI:ele_sg_dist_high}}, some screened high-$\kappa_{\mathrm{L}}$ candidates also belong to lower-symmetry space groups, such as BeCN$_2$ ($I\overline{4}2d$, No. 122), MoO$_2$ ($R3m$, No. 160) and B$_6$O ($R\overline{3}m$, No. 166).

Because complete DFT-based anharmonic thermal-transport calculations for all 463 low-$\kappa_{\mathrm{L}}$ candidates, including three- and four-phonon scattering, second-order force-constant renormalization, and coherent contributions, would be prohibitively expensive, we applied a second-stage descriptor-based selection to reduce the number of candidates for DFT validation. The detailed selection criteria are summarized in Table~\textcolor{darkteal}{S2}\ignore{\ref{tab:secondary_low_kappa_screening}}. Following this refinement, 253 low-$\kappa_{\mathrm{L}}$ candidates were retained for DFT-based validation. Because the screening criteria favor materials with large MSD values and soft lattice dynamics, a substantial fraction of these candidates exhibited significant imaginary phonon modes in the subsequent DFT calculations. Such modes preclude reliable evaluation of $\kappa_{\mathrm{L}}$. In addition, calculations for several compounds could not be completed because of failures in extracting higher-order interatomic force constants, numerical issues, or convergence difficulties during the ShengBTE calculations. Consequently, $\kappa_{\mathrm{L}}$ calculations were completed successfully for 149 low-$\kappa_{\mathrm{L}}$ candidates. The corresponding results are summarized in Tables~\textcolor{darkteal}{S5} and~\textcolor{darkteal}{S6} of the Supplementary Information.

For the high-$\kappa_{\mathrm{L}}$ screening, 49 preliminary candidates were identified using the criteria MSD < 0.02~\AA$^2$ and $\omega_{\min}^{\mathrm{opt}}$ > 10 THz. Several structures were subsequently found to exhibit significant imaginary phonon modes at the DFT level, preventing reliable calculation of their $\kappa_{\mathrm{L}}$. Consequently, accurate $\kappa_{\mathrm{L}}$ values were obtained for 34 compounds. The corresponding results are summarized in Tables~\textcolor{darkteal}{S3} and~\textcolor{darkteal}{S4} of the Supplementary Information.

To assess the reliability of our calculated $\kappa_{\mathrm{L}}$, we compared our results with the corresponding values reported in the Phonix~\cite{WOS:001739467100001} database and in the high-throughput $\kappa_{\mathrm{L}}$ calculations by Li et al.~\cite{Li2025}. Our dataset shares 55 compounds with the Phonix~\cite{WOS:001739467100001} database and 29 with the dataset of Li et al.~\cite{Li2025}. As shown in Fig.~\ref{fig:ltc_compare}, both comparisons show excellent agreement, with $R^2$ values exceeding 0.95. Nevertheless, relatively large deviations occur for several materials in the high-$\kappa_{\mathrm{L}}$ regime, including BN, BAs, BP, SiC, and GaN in the $F\bar{4}3m$ space group. These discrepancies may arise from differences in the treatment of phonon scattering and higher-order anharmonicity, as well as from differences in computational methods and parameter settings. For example, Phonix~\cite{WOS:001739467100001} uses the \textsc{ALAMODE} package~\cite{Tadano_2014} to calculate $\kappa_{\mathrm{L}}^{\mathrm{P}}$ and $\kappa_{\mathrm{L}}^{\mathrm{C}}$, but considers only three-phonon scattering, whereas our $\kappa_{\mathrm{L}}^{\mathrm{P}}$ calculations include both three- and four-phonon scattering with the \textsc{FourPhonon} package~\cite{HAN2022108179}. Although Li et al.~\cite{Li2025} also used the \textsc{FourPhonon} package~\cite{HAN2022108179}, their computational parameters differ from ours. For instance, they used a fixed $q$-mesh density for all phonon Boltzmann transport equation calculations, including those involving four-phonon scattering, whereas we performed $q$-mesh convergence tests for each material individually. In addition, their supercells contained approximately 120 atoms on average, whereas ours typically contained 120--300 atoms. These differences may introduce relative errors in the calculated $\kappa_{\mathrm{L}}$. For materials with intrinsically high $\kappa_{\mathrm{L}}$, even moderate relative differences can result in substantial absolute deviations.

We then benchmarked our calculated $\kappa_{\mathrm{L}}$ against experimental data. Four compounds exhibit relatively large deviations: BN, BAs, BP, and SiC in the $F\bar{4}3m$ structure, with experimental values of 880, 1000, 460, and 500 Wm$^{-1}$K$^{-1}$, respectively~\cite{10.1126/science.aaz6149,10.1126/science.aat8982,10.1021/acs.nanolett.7b03437,Cheng2022}. Our corresponding calculated values are 911, 1189, 497, and 490 Wm$^{-1}$K$^{-1}$, differing from the experimental values by approximately 3.6\%, 18.9\%, 8.0\%, and 2.4\%, respectively. These deviations are smaller than those reported in the Phonix~\cite{WOS:001739467100001} database for BN (679 Wm$^{-1}$K$^{-1}$) and BAs (1344 Wm$^{-1}$K$^{-1}$), and those reported by Li et al.~\cite{Li2025} for BP (615 Wm$^{-1}$K$^{-1}$) and SiC (574 Wm$^{-1}$K$^{-1}$). Overall, our room-temperature values agree well with experiment. Our calculated $\kappa_{\mathrm{L}}$ for cubic GaN (195 Wm$^{-1}$K$^{-1}$ at 300 K) falls between the values of 181 and 215 Wm$^{-1}$K$^{-1}$ reported in previous first-principles calculations~\cite{TANG2021121659,PhysRevLett.109.095901}. These comparisons support the reliability of our calculations.

Figures~\ref{fig:ltc_distribution}(a) and (b) summarize the distributions of the low-$\kappa_{\mathrm{L}}$ and high-$\kappa_{\mathrm{L}}$ datasets, respectively. In the low-$\kappa_{\mathrm{L}}$ dataset, nearly all screened materials have $\kappa_{\mathrm{L}}$ below 2 Wm$^{-1}$K$^{-1}$, and approximately 70\% have $\kappa_{\mathrm{L}} \leq 1$ Wm$^{-1}$K$^{-1}$. This distribution demonstrates the effectiveness of the combined criterion of large MSD and low $\omega_{\min}^{\mathrm{opt}}$ for identifying materials with low $\kappa_{\mathrm{L}}$. In the high-$\kappa_{\mathrm{L}}$ dataset, approximately 44\% of the screened materials have $\kappa_{\mathrm{L}}$ above 100 Wm$^{-1}$K$^{-1}$. Although this success rate is lower than that for low-$\kappa_{\mathrm{L}}$ materials, it still indicates substantial enrichment in high-$\kappa_{\mathrm{L}}$ materials relative to the unfiltered dataset. Overall, these results demonstrate that MSD and $\omega_{\min}^{\mathrm{opt}}$ provide an effective descriptor-based strategy for screening materials with extreme $\kappa_{\mathrm{L}}$ values.

\begin{figure*}[tph!]
	\includegraphics[clip,width=1.0\linewidth]{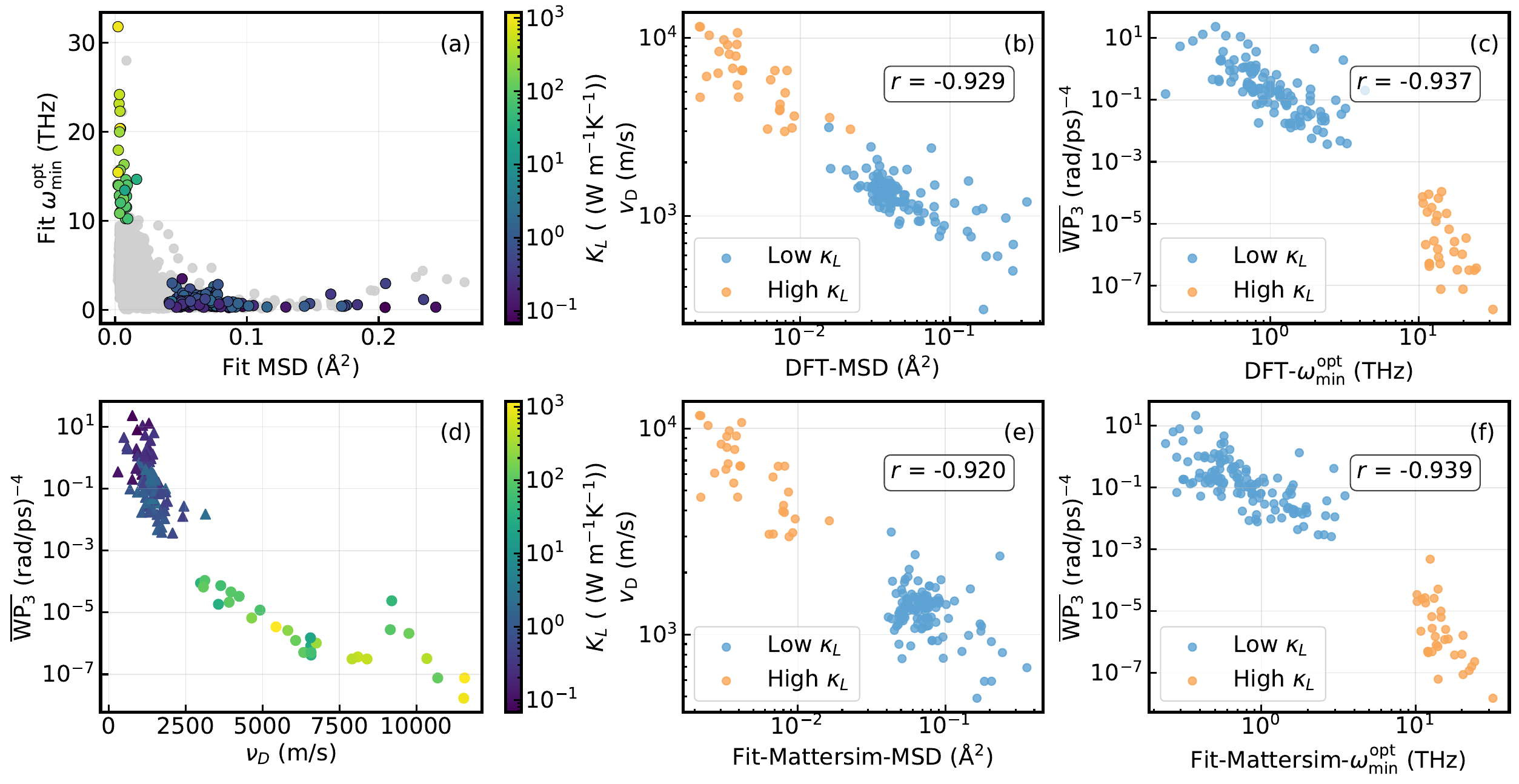}
	\caption{(a) Correlation between MSD and $\omega_{\min}^{\mathrm{opt}}$ predicted by the tuned Fit‑Mattersim potential versus DFT‑calculated lattice thermal conductivity. (d) Correlation among three‑phonon scattering phase space $\overline{\mathrm{WP_3}}$, Debye sound velocity($\nu_{D}$), and DFT‑calculated lattice thermal conductivity. Panels (b,c) are derived from DFT calculations at 300\,K: (b) Scatter plot showing the correlation between DFT‑calculated mean squared displacement (MSD) and Debye sound velocity($\nu_{D}$); (c) Correlation between DFT‑computed minimum optical frequency $\omega_\mathrm{min}$ and average weighted three-phonon scattering phase space $\overline{\mathrm{WP_3}}$. Panels (e,f) adopt the same descriptor correlation forms as (b,c), where the MSD and $\omega_\mathrm{min}$ values are predicted by the tuned Fit‑Mattersim machine‑learning potential. Blue and orange dots represent low‑$\kappa_{\mathrm{L}}$ and high‑$\kappa_{\mathrm{L}}$ materials, respectively. The Pearson correlation coefficient $r$ for each logarithmic‑scale relationship is annotated in the upper‑right corner of each subpanel.}
	\label{fig:kappa_anylze}
\end{figure*}

Having established the reliability of the calculated $\kappa_{\mathrm{L}}$, we next evaluated the effectiveness of the descriptor-based screening strategy. Figure~\ref{fig:kappa_anylze}(a) shows the distribution of the validated materials in the MSD--$\omega_{\min}^{\mathrm{opt}}$ descriptor space. As anticipated, all validated materials within the designated low-$\kappa_{\mathrm{L}}$ screening region, characterized by large MSD and low $\omega_{\min}^{\mathrm{opt}}$, exhibit low $\kappa_{\mathrm{L}}$. Conversely, materials within the designated high-$\kappa_{\mathrm{L}}$ screening region, characterized by small MSD and high $\omega_{\min}^{\mathrm{opt}}$, exhibit high $\kappa_{\mathrm{L}}$.

To further elucidate the origins of the low and high $\kappa_{\mathrm{L}}$ values among the screened materials, we analyzed the relationships among the Debye sound velocity $\nu_{\mathrm{D}}$ ($\nu_{\mathrm{D}}$ denotes the average acoustic velocity estimated from the elastic moduli and mass density~\cite{CHEN2001947,huang2022inner,ledbetter2001monocrystal,ANDERSON1963909,naher2021ab}; see Section 3 of the SI), the average weighted three-phonon scattering phase space $\overline{\mathrm{WP_3}}$ (the mode-resolved weighted phase space averaged over all phonon branches and the full Brillouin zone~\cite{PhysRevB.91.144304,PhysRevB.110.104308,10.1039/d2nh00429a}; see Section 3 of the SI), and $\kappa_{\mathrm{L}}$, as shown in Fig.~\ref{fig:kappa_anylze}(d). The screened low-$\kappa_{\mathrm{L}}$ materials are concentrated in the low-$\nu_{\mathrm{D}}$, high-$\overline{\mathrm{WP_3}}$ region, with most exhibiting $\nu_{\mathrm{D}}$ < 3200 ms$^{-1}$ and $\overline{\mathrm{WP_3}}$ above 3.5 $\times$ 10$^{-3}$ ps$^4$rad$^{-4}$. Materials with $\kappa_{\mathrm{L}}$ < 1 Wm$^{-1}$K$^{-1}$ are further concentrated at $\nu_{\mathrm{D}}$ < 2000 ms$^{-1}$ and particularly high $\overline{\mathrm{WP_3}}$ values. These observations are consistent with slow acoustic-phonon propagation and a large weighted phase space that favors three-phonon scattering. By contrast, the screened high-$\kappa_{\mathrm{L}}$ materials span broader ranges of $\nu_{\mathrm{D}}$ and $\overline{\mathrm{WP_3}}$, consistent with their $\kappa_{\mathrm{L}}$ values ranging from 11 to 1189 Wm$^{-1}$K$^{-1}$. Nevertheless, the candidates with the highest $\kappa_{\mathrm{L}}$ values are concentrated in the high-$\nu_{\mathrm{D}}$, low-$\overline{\mathrm{WP_3}}$ region, generally exhibiting $\nu_{\mathrm{D}}$ > 7900 ms$^{-1}$ and $\overline{\mathrm{WP_3}}$ < 2.4 $\times$ 10$^{-5}$ ps$^4$rad$^{-4}$. Their high $\kappa_{\mathrm{L}}$ values are therefore consistent with rapid acoustic-phonon propagation and a small weighted phase space for three-phonon scattering.

To clarify how the MSD--$\omega_{\min}^{\mathrm{opt}}$ descriptors capture these phonon-transport properties, we quantified the correlations between MSD and $\nu_{\mathrm{D}}$ and between $\omega_{\min}^{\mathrm{opt}}$ and $\overline{\mathrm{WP_3}}$, as shown in Fig.~\ref{fig:kappa_anylze}(b) and (c). All Pearson correlation coefficients reported below were calculated using logarithmically transformed values of both variables. The DFT-calculated MSD and $\nu_{\mathrm{D}}$ exhibit a strong negative correlation, with a coefficient of -0.929, indicating that large atomic vibrational amplitudes are generally associated with soft lattices and low $\nu_{\mathrm{D}}$. Similarly, the DFT-calculated $\omega_{\min}^{\mathrm{opt}}$ and $\overline{\mathrm{WP_3}}$ are strongly negatively correlated, with a coefficient of -0.937, indicating that low-frequency optical modes are associated with a larger $\overline{\mathrm{WP_3}}$. These relationships are preserved when MSD and $\omega_{\min}^{\mathrm{opt}}$ are calculated using Fit-MatterSim, with correlation coefficients of -0.920 and -0.939 with the DFT-calculated $\nu_{\mathrm{D}}$ and $\overline{\mathrm{WP_3}}$, respectively, as shown in Fig.~\ref{fig:kappa_anylze}(e) and (f). Thus, large MSD and low $\omega_{\min}^{\mathrm{opt}}$ preferentially select materials with low $\nu_{\mathrm{D}}$ and large $\overline{\mathrm{WP_3}}$, whereas small MSD and high $\omega_{\min}^{\mathrm{opt}}$ favor the opposite characteristics. Together, these descriptors capture complementary aspects of phonon propagation and scattering that are relevant to screening materials with low or high $\kappa_{\mathrm{L}}$.

\begin{figure*}[tph!]
	\includegraphics[clip,width=1\linewidth]{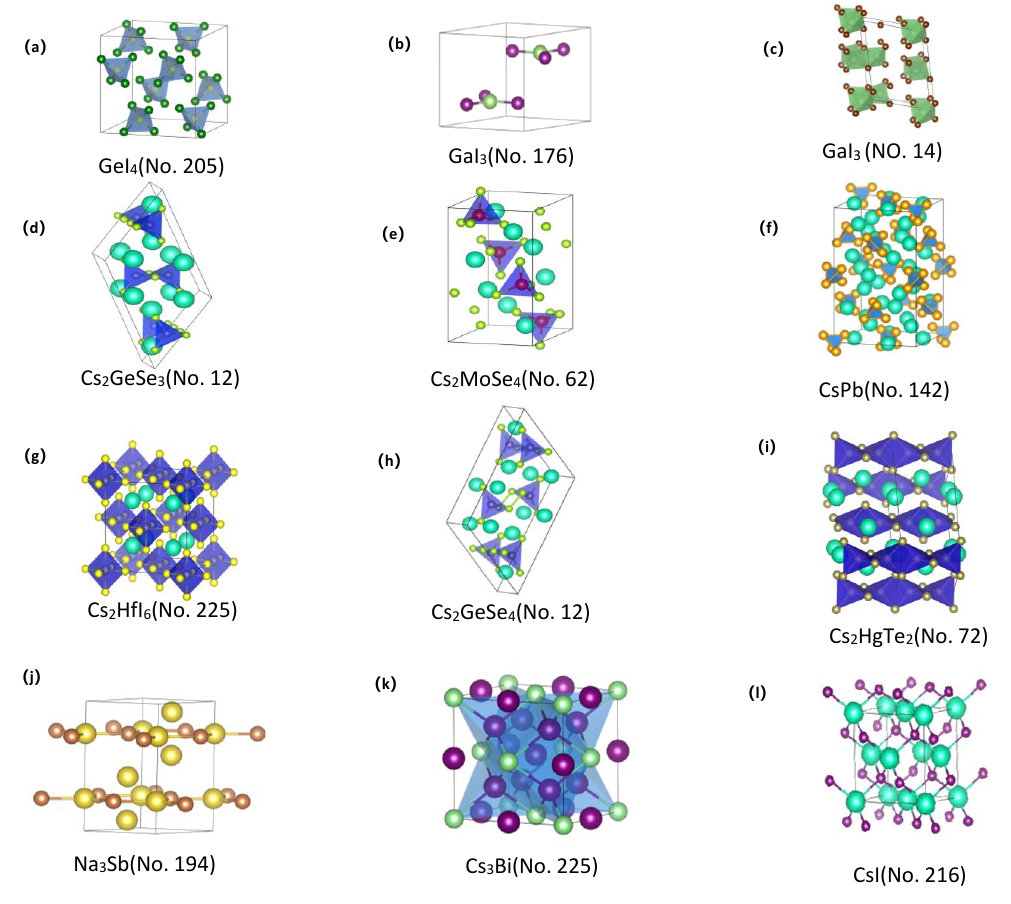}
	\caption{Crystal‑structure classification of investigated materials. Panels (a)–(c) show representative examples of molecular crystals or molecular‑crystal‑like structures; (d)–(h) illustrate heavy‑cation‑separated discrete anionic structures; (i)–(k) correspond to heavy‑element‑filled host–guest‑like structures; and (l) depicts a typical weakly bonded continuous‑network structure.}
	\label{fig:stru}
\end{figure*}

\begin{table*}[tph!]
	\centering
	\normalsize
	\captionsetup{justification=centering}
	\fontsize{10}{12}\selectfont
	\caption{Representative structures of candidates with calculated lattice thermal conductivities $\kappa_\mathrm{L}<1~\mathrm{W\cdot m^{-1}\cdot K^{-1}}$. $N_\mathrm{prim}$ denotes the number of atoms in the primitive cell. Stru.Type indicates the structure type assigned in the ICSD; for materials not listed in the ICSD, their chemical formulas are used instead. Number of Compound denotes the number of materials calculated in this study belonging to each structure type. Only selected representative materials are shown here; the complete list of 104 materials with $\kappa_\mathrm{L}<1~\mathrm{W\cdot m^{-1}\cdot K^{-1}}$ is provided in Table \textcolor{darkteal}{S5}.}
	\renewcommand{\arraystretch}{1.2}
	\renewcommand{\theadfont}{}
	\setlength{\tabcolsep}{2pt}
	\label{tab:lower_ltc}
	\begin{tabular}{cccccccccc}
		\toprule
		Compound & OQMD\_ID  & ICSD\_ID & Stru.Type & \makecell{Number of \\ Compound} &  S.G. & $N_{\mathrm{prim}}$ & \makecell{$\Delta E_{\text{hull}}$ \\\footnotesize($\mathrm{eV/atom)}$}  & \makecell{Bandgap \\\footnotesize(eV)}  & \makecell{$\kappa^{xx}_L$,$\kappa^{yy}_L$, $\kappa^{zz}_L$ \\[3pt]\footnotesize(Wm$^{-1}$K$^{-1}$)}  \\
		\midrule
		HfI$_4$            & 19110  & 109275 & SnI$_4$              & 7 & 205 & 40 &  0.03 & 3.12  & 0.13,0.13,0.13  \\
		GaI$_3$            &25812   & 413457 & AlI$_3$              & 5 &  14  & 16 & 0.00 & 2.11 & 0.15,0.16,0.19  \\
		CsRb$_2$Sb         &559719  &        & Cu$_2$MnAl           & 9 &  225 & 4  & 0.02  & 0.83 & 0.53,0.51,0.53  \\
		Rb$_3$Bi           &10961   & 58849  & Cu$_2$MnAl           & 9 &  225 & 4  & 0.03  & 0.08 & 0.38,0.38,0.38  \\
		Cs$_2$GeSe$_4$     &13422   & 74832  & Cs$_2$GeSe$_4$       & 3 &  12  & 14 & 0.00 & 1.53 & 0.34,0.39,0.35  \\
		Cs$_5$Te$_3$       &6410    & 34000  & Cs$_5$Te$_3$         & 1 &  12  & 16 & 0.02  & 0.41 & 0.30,0.34,0.28  \\
		Cs$_2$Se           &2026374 & 627099 & PbCl$_2$             & 4 &  62  & 12 & 0.02  & 1.93 & 0.44,0.44,0.49  \\
		CsSe               &1361113 &        & CsSe                 & 1 &  142 & 16 & 0.02  & 1.77 & 0.26,0.26,0.30  \\
		Cs$_4$PbBr$_6$     &20521   & 162158 & K$_4$CdCl$_6$        & 7 &  167 & 22 & 0.00  & 3.78 & 0.23,0.23,0.28  \\
		Cs$_3$SnTe$_3$     &24133   & 280070 & Cs$_6$Ge$_2$Te$_6$   & 4 &  15  & 28 & 0.00 & 1.12 & 0.32,0.19,0.23  \\
		CsGe               &29603   & 626996 & NaPb                 & 7 &  142 & 32 & 0.02  & 0.89 & 0.41,0.41,0.35  \\
		Cs$_3$InTe$_3$     &1975575 &        & Cs$_3$InTe$_3$       & 1 &  12  & 14 & 0.00 & 1.80 & 0.40,0.24,0.38  \\
		Cs$_2$MoSe$_4$     &29606   & 627043 & $\beta$-K$_2$SO$_4$  & 4 &  62  & 28 & 0.00 & 1.11 & 0.24,0.55,0.28  \\
		Na$_3$Sb           &5156    & 26882  & Na$_3$As             & 5 &  194 & 8  & 0.00 & 0.42 & 0.63,0.63,0.53  \\
		Tl$_2$Te$_3$Pb$_2$ &1477605 &        & Tl$_2$Te$_3$Pb$_2$   & 1 &  15  & 14 & 0.07  & 0.82 & 0.43,0.41,0.52  \\
		Cs$_2$HgTe$_2$     &1745180 &        & Cs$_2$HgTe$_2$       & 1 &  72  & 10 & 0.00  & 1.52 & 0.36,0.39,0.58  \\
		BiI                &1952    & 1559   & BiI                  & 1 &  12  & 8  & 0.03  & 0.57 & 0.39,1.48,0.30  \\
		\bottomrule
	\end{tabular}
\end{table*}

\subsection{Selected High and Ultralow Thermal Conductivity Materials}\label{section5}
We calculated $\kappa_{\mathrm{L}}$ for 34 of the 49 high-$\kappa_{\mathrm{L}}$ candidates identified in the initial screening. For the remaining 15 candidates, imaginary phonon frequencies, difficulties in extracting higher-order interatomic force constants, or convergence problems in the \textsc{ShengBTE} calculations prevented us from obtaining $\kappa_{\mathrm{L}}$ values. Of the 34 candidates with completed calculations, 15 exhibit room-temperature $\kappa_{\mathrm{L}}$ values above $100~\mathrm{W\,m^{-1}\,K^{-1}}$, including 10 with values above $200~\mathrm{W\,m^{-1}\,K^{-1}}$ (Table~\textcolor{darkteal}{S3}).

Materials with high $\kappa_{\mathrm{L}}$ are less common than those with low $\kappa_{\mathrm{L}}$. High $\kappa_{\mathrm{L}}$ generally favors a combination of strong chemical bonds, a simple crystal structure, and light constituent elements, which limits the number of suitable candidates. Many of the materials listed in Table~\textcolor{darkteal}{S3}, particularly those with $\kappa_{\mathrm{L}}$ above $200~\mathrm{Wm^{-1}K^{-1}}$, have previously been shown experimentally or predicted theoretically to exhibit high $\kappa_{\mathrm{L}}$~\cite{PhysRevB.87.165201,10.1002/adfm.201805116,10.1126/science.aat8982,10.1126/science.aaz6149,PhysRevMaterials.5.053801,10.1063/5.0034906,10.1063/1.4827419,PhysRevB.88.144306}. Although the screening did not identify previously unexplored high-$\kappa_{\mathrm{L}}$ materials, it recovered well-established examples, including BAs, BN, TaN, BP, SiC, and BeO. This result supports the use of the selected descriptors to identify materials with physical characteristics favorable to high $\kappa_{\mathrm{L}}$.

Of the 149 low-$\kappa_{\mathrm{L}}$ candidates for which the calculations were completed, 104 have $\kappa_{\mathrm{L}}$ < 1~$\mathrm{Wm^{-1}K^{-1}}$. Of these 104 candidates, 65 have experimentally reported crystal structures. Table~\ref{tab:lower_ltc} lists representative structure types for which at least one compound has a calculated $\kappa_{\mathrm{L}}$ below this threshold. The complete list is provided in Table~\textcolor{darkteal}{S5} of the Supporting Information. The close agreement between calculated and experimental lattice parameters for representative low-$\kappa_{\mathrm{L}}$ compounds (\textcolor{darkteal}{Table~S7}\ignore{Table~\ref{Tab_SI:lattice}}) supports the suitability of the computational settings used here. Based on their structural characteristics and the root-mean-square (RMS) values of their interatomic force constants, we classify the materials with $\kappa_{\mathrm{L}} < 1~\mathrm{W\,m^{-1}\,K^{-1}}$ into four categories:

\begin{enumerate}[label=\roman*., leftmargin=*]
	\item \textbf{Molecular crystals and structures resembling molecular crystals}: These structures contain identifiable neutral molecular units. Strong covalent bonds connect atoms within each unit, whereas van der Waals forces or other weak intermolecular interactions bind neighboring units~\cite{Batsanov:me6008,Shiga2026}. Representative prototypes include GeI$_4$, GaI$_3$, and GaBr$_3$.
	
	\item \textbf{Structures with discrete anionic units separated by heavy cations}: Unlike the neutral molecules in the first category, the discrete units in these structures are charged inorganic motifs, such as anionic clusters or isolated polyhedra. Heavy cations such as Cs$^+$ and Rb$^+$ occupy the spaces between these units and provide charge balance~\cite{Jung2021,10.1039/d2cc00258b}. Representative prototypes include  Cs$_2$MoSe$_4$, Cs$_2$HfI$_6$, and CsPb.
	
	\item \textbf{Host--guest-like structures containing heavy elements}: In these structures, the host atoms form extended chains, layers, or frameworks. Heavy atoms such as Cs and Rb occupy the spaces between chains or layers or the voids within frameworks, giving rise to a host--guest-like arrangement~\cite{RevModPhys.86.669,PhysRevLett.114.095501,Ikeda2019,Christensen2008}. Representative prototypes include Cs$_2$HgTe$_2$, Na$_3$Sb, and Cs$_3$Bi.
	
	\item \textbf{Weakly bonded continuous networks}: These structures lack discrete molecular groups or isolated polyatomic motifs. Instead, relatively weak bonds connect the constituent atoms throughout the crystal, resulting in networks with high connectivity despite the softness of the bonds~\cite{10.1002/adfm.202108532,PhysRevB.108.174302}. Representative prototypes include CsI and RbI.
\end{enumerate}

Across the full range of materials considered, differences in bond strength and atomic mass contribute substantially to the variation in $\kappa_{\mathrm{L}}$. The Debye velocity, $\nu_{\mathrm{D}}$, calculated from the elastic constants and mass density, reflects the combined effects of these factors and correlates with $\kappa_{\mathrm{L}}$ over this broad range (Fig.~\ref{fig:kappa_anylze}). However, optical phonons and acoustic phonons away from the $\Gamma$ point of the Brillouin zone can also contribute appreciably to heat transport. Because $\nu_{\mathrm{D}}$ describes the long-wavelength acoustic limit, it does not fully capture these contributions. We therefore also considered the average phonon group velocity, $\langle \nu \rangle$, a descriptor used in previous studies of lattice thermal transport~\cite{10.3389/fenrg.2018.00034,10.1063/5.0155409,10.1063/1.4979871,10.1039/d0cp03191g,Ma2019}. The results are presented in Section~2 of the Supporting Information.

For materials with $\kappa_{\mathrm{L}}$ < 1~$\mathrm{Wm^{-1}K^{-1}}$, the Pearson correlation coefficient between $\nu_{\mathrm{D}}$ and $\kappa_{\mathrm{L}}$ is 0.398 (Fig.~\textcolor{darkteal}{S6}\ignore{\ref{fig_SI:velo_ltc_velocity}}). By comparison, the coefficient for $\langle \nu \rangle$ and $\kappa_{\mathrm{L}}$ is 0.775 (Fig.~\ref{fig:velo_ltc}). This stronger correlation highlights the importance of accounting for phonon propagation beyond the long-wavelength acoustic limit in materials with ultralow $\kappa_{\mathrm{L}}$.

The compounds marked by stars in Fig.~\ref{fig:velo_ltc} fall below the overall trend: their $\kappa_{\mathrm{L}}$ values are substantially lower than those of other compounds with comparable $\langle \nu \rangle$. As discussed below, stronger phonon scattering further suppresses heat transport in these compounds. The compounds marked by circles exhibit weaker phonon scattering. For this subset, the Pearson correlation coefficient between $\langle \nu \rangle$ and $\kappa_{\mathrm{L}}$ is 0.899, suggesting that differences in phonon group velocity largely account for the variation in $\kappa_{\mathrm{L}}$. In the following sections, we examine the distributions of $\langle \nu \rangle$ and $\kappa_{\mathrm{L}}$ within each structural category and use representative compounds to illustrate the physical origins of ultralow $\kappa_{\mathrm{L}}$.

\subsubsection{Molecular crystals or molecular-crystal-like structures}
Figure~\ref{fig:stru}(a--c) shows representative compounds in this category. The connectivity of their local structural motifs distinguishes two types: isolated monomers and discrete dimers formed by edge-sharing tetrahedra. GeI$_4$ ($Pa\overline{3}$, No.~205) and GaI$_3$ ($P6_3/m$, No.~176), shown in Fig.~\ref{fig:stru}(a) and (b), respectively, exemplify the first type. In GeI$_4$, each molecule consists of a Ge atom tetrahedrally coordinated by four I atoms. The average root-mean-square (RMS) value of the second-order interatomic force constants ($\Phi_{ij}^{(2)}$), $\langle \mathrm{RMS}(\Phi_{ij}^{(2)}) \rangle$, is $2.595~\mathrm{eV}/\text{\AA}^{2}$ for the four intramolecular Ge--I interactions, compared with only 0.06 eV/\AA$^{2}$ for the nearest intermolecular I$\cdots$I contact. In GaI$_3$, each molecule consists of one Ga atom and three I atoms in a trigonal planar arrangement. The corresponding values are 2.98 eV/\AA$^{2}$ for the intramolecular Ga--I interactions and 0.038 eV/\AA$^{2}$ for the nearest intermolecular I$\cdots$I contact. Thus, despite their different geometries, both crystals consist of strongly bonded molecules with much weaker interactions between them.

The monoclinic GaI$_3$ phase ($P2_1/c$, No.~14) exemplifies the second type [Fig.~\ref{fig:stru}(c)]. Two Ga-centered tetrahedra share an edge to form a discrete dimer. The bridging Ga--I interactions within each dimer have an average RMS force constant of 1.29 eV/\AA$^{2}$, compared with only 0.06 eV/\AA$^{2}$ for the nearest I$\cdots$I contact between dimers. Although the local geometry is more complex than in the monomeric structures, strong bonding remains confined to discrete units, preserving the zero-dimensional character of the bonding network.

The neutral monomers and dimers interact primarily through van der Waals forces. This weak coupling produces small restoring forces for relative translations and librations of neighboring units, giving rise to low-frequency optical modes and large mean-square displacements (MSDs) at a given temperature. Both the low minimum optical phonon frequency, $\omega_{\min}^{\mathrm{opt}}$, and the large MSDs are consistent with the screening criteria. Weak coupling between units also gives rise to weakly dispersive phonon branches and reduces the slopes of the acoustic branches near the $\Gamma$ point, thereby limiting phonon group velocities. As shown in Fig.~\ref{fig:velo_ltc}, molecular crystals and related structures generally exhibit the lowest $\langle \nu \rangle$ and $\kappa_{\mathrm{L}}$ of the four categories. All compounds in this category have $\kappa_{\mathrm{L}}$ < 0.2 Wm$^{-1}$K$^{-1}$ at room temperature. These results indicate that, in the absence of a continuous three-dimensional framework of strong bonds, weak coupling between structural units limits phonon propagation and contributes to ultralow $\kappa_{\mathrm{L}}$.

The two BBr$_3$ phases ($P2_1/c$, No.~14, and $P6_3/m$, No.~176) have relatively high $\langle \nu \rangle$ within this category, yet both retain ultralow $\kappa_{\mathrm{L}}$. Their average atomic masses are relatively low (Fig.~\textcolor{darkteal}{S7}\ignore{\ref{fig_SI:velo_ltc_avg_mass}}), whereas the RMS values of the second-order interatomic force constants indicate that interactions between neighboring units are comparable in the BBr$_3$ and GaI$_3$ phases of the same space group (Fig.~\textcolor{darkteal}{S8}\ignore{\ref{fig_SI:fc2_rms_bar}}). This comparison suggests that the higher group velocities in BBr$_3$ arise primarily from its lower average atomic mass rather than from stronger coupling between units.

Figure~\textcolor{darkteal}{S9}\ignore{\ref{fig_SI:TYPE_1_WP}} compares the phonon spectra, mode Gr\"uneisen parameters, and phonon scattering rates of these structures. In the low-frequency range, both BBr$_3$ phases exhibit larger absolute values of the mode Gr\"uneisen parameters than GaI$_3$ ($P6_3/m$, No.~176). This indicates stronger anharmonicity and is consistent with the higher scattering rates of the BBr$_3$ phases in the same frequency range. The resulting shorter phonon lifetimes explain why $\kappa_{\mathrm{L}}$ remains exceptionally low in both phases despite their relatively high average phonon group velocities.

\subsubsection{Heavy-cation-separated discrete anionic structures}
Figure~\ref{fig:stru}(d--h) shows representative structures in this category. As in molecular crystals, these structures contain discrete structural units. Here, the units are charged inorganic motifs, such as anionic clusters or isolated polyhedra. Heavy alkali-metal cations, typically Cs$^+$ or Rb$^+$, separate the anionic units and provide charge balance. Electrostatic interactions between the cations and anionic units thus contribute to crystal cohesion. In molecular crystals, cohesion between neutral units arises primarily from van der Waals interactions.

\begin{figure}[tph!]
	\includegraphics[clip,width=1\linewidth]{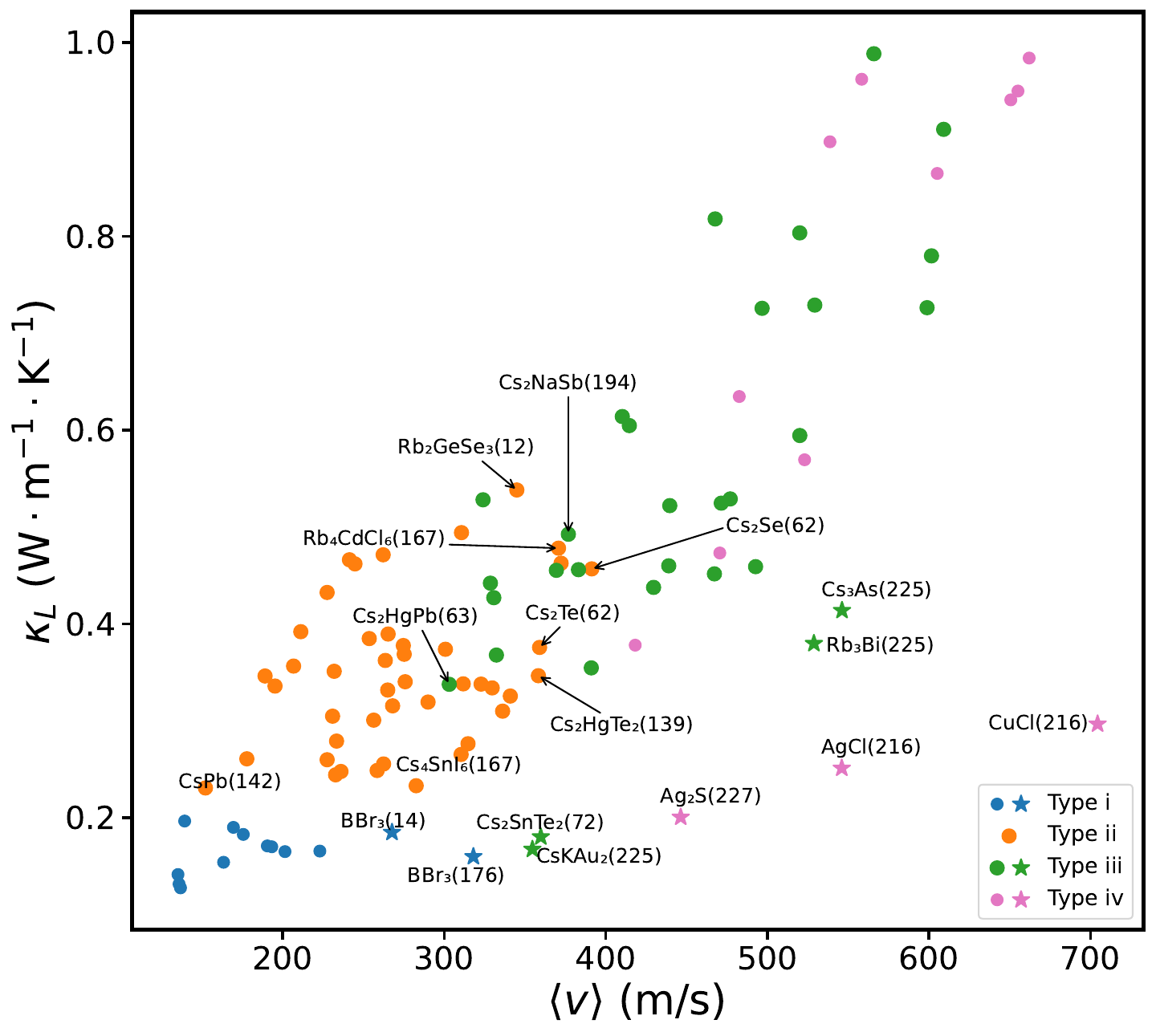}
	\caption{Correlation between $\langle v \rangle$ and lattice thermal conductivity. Each data point is colored according to the crystal-structure classification: blue, molecular crystals or molecular-crystal-like structures; orange, heavy-cation-separated discrete anionic structures; green, heavy-element-filled host–guest-like structures; pink, weakly bonded continuous-network structures. Star-shaped markers denote structures with strong phonon scattering, which deviate substantially from the overall trend. The Pearson correlation coefficient reaches 0.775 when all structures are considered, and increases to 0.899 after excluding these strongly scattering structures (star-shaped markers).}
	\label{fig:velo_ltc}
\end{figure}

To quantify the coupling between the heavy cations and the surrounding anionic units, we calculated $\langle \mathrm{RMS}(\Phi_{ij}^{(2)}) \rangle$ for each heavy cation and its nearest neighbors (Fig.~\textcolor{darkteal}{S10}\ignore{~\ref{fig_SI:rms}}). For nearest-neighbor Cs--$X$ and Rb--$X$ pairs, these values range from 0.058 to 0.21~eV/\AA$^{2}$. They are generally larger than the corresponding values for interactions between neighboring molecular units in the first category, consistent with stronger coupling mediated by electrostatic interactions. Materials in this category also generally exhibit higher Debye velocities, $\nu_{\mathrm{D}}$, consistent with relatively stiff lattices (Fig.~\textcolor{darkteal}{S6}\ignore{\ref{fig_SI:velo_ltc_velocity}}). Their average phonon group velocities, $\langle \nu \rangle$, likewise tend to exceed those of molecular crystals and related structures (Fig.~\ref{fig:velo_ltc}).

The cation--anion force constants nevertheless remain relatively small. The resulting weak restoring forces favor large mean-square displacements of the heavy cations and, together with the large cation masses, give rise to low-frequency optical modes and a small $\omega_{\min}^{\mathrm{opt}}$. Phonon propagation is limited by the relatively weak cation--anion coupling and the absence of a continuous three-dimensional network of strong bonds. Low-frequency modes involving the heavy cations can also be localized or weakly dispersive and may enhance the scattering of heat-carrying acoustic phonons. Although the stronger coupling between structural units generally gives rise to higher $\kappa_{\mathrm{L}}$ than in molecular crystals and related structures, most materials in this category still exhibit ultralow $\kappa_{\mathrm{L}}$, with values between 0.2 and 0.4~Wm$^{-1}$K$^{-1}$ (Fig.~\ref{fig:velo_ltc}).

Within this category, CsPb ($I4_1/acd$, No.~142) has an exceptionally low $\langle \nu \rangle$, comparable to values found in molecular-crystal-like structures. To examine the origin of this low velocity, we compared CsPb with the isostructural compounds CsGe and CsSn. The Cs--Ge, Cs--Sn, and Cs--Pb interactions have similar $\langle \mathrm{RMS}(\Phi_{ij}^{(2)}) \rangle$ values, indicating comparable cation--anion interaction strengths (Fig.~\textcolor{darkteal}{S11}\ignore{\ref{fig_SI:CsSb_ifc_rms}}). Atomic mass also influences phonon propagation through the mass-weighting factor $1/\sqrt{m_i m_j}$ in the dynamical matrix. For comparable force constants, larger atomic masses generally favor lower phonon group velocities. CsPb has the largest average atomic mass in this structural category (Fig.~\textcolor{darkteal}{S7}\ignore{\ref{fig_SI:velo_ltc_avg_mass}}). Given the similar cation--anion force constants across the three compounds, the large average atomic mass of CsPb likely further limits phonon propagation in a structure that already lacks a continuous network of strong bonds. Its low $\langle \nu \rangle$ thus contributes to its low $\kappa_{\mathrm{L}}$.

Rb$_2$GeSe$_3$ ($C2/m$, No.~12) has the highest directionally averaged $\kappa_{\mathrm{L}}$ in this category, although other compounds have higher $\langle \nu \rangle$. Its relatively high average conductivity arises from enhanced heat transport along the $y$ direction: $\kappa_{\mathrm{L}}$ reaches approximately 0.75 Wm$^{-1}$K$^{-1}$ along $y$, compared with 0.44 and 0.42 Wm$^{-1}$K$^{-1}$ along $x$ and $z$, respectively. The phonon group velocities are also higher along $y$ than along $x$ or $z$ (Fig.~\textcolor{darkteal}{S12}\ignore{\ref{fig_SI:Rb2GeSe3}}). To investigate the origin of this directional dependence, we examined the elastic stiffness tensor of Rb$_2$GeSe$_3$. Its $C_{22}$ is the largest in this category, whereas its $C_{11}$ and $C_{33}$ are comparable to those of other materials in the category (Fig.~\textcolor{darkteal}{S13}\ignore{\ref{fig_SI:Rb2GeSe3_Cxx}}). The large $C_{22}$ indicates high longitudinal stiffness along $y$, consistent with the higher phonon group velocities and enhanced heat transport in that direction. The scalar average $\langle \nu \rangle$ obscures this directional dependence and therefore only partly accounts for the relatively high average $\kappa_{\mathrm{L}}$ of Rb$_2$GeSe$_3$.

Cs$_2$Se ($Pnma$, No.~62) exhibits the highest $\langle \nu \rangle$ in this category. It also has the largest cation--anion $\langle \mathrm{RMS}(\Phi_{ij}^{(2)}) \rangle$, indicating relatively strong Cs--Se interactions (Fig.~\textcolor{darkteal}{S10}\ignore{\ref{fig_SI:rms}}). Comparison with isostructural Cs$_2$Te ($Pnma$, No.~62) illustrates the additional influence of atomic mass. The two compounds have comparable cation--anion force constants (Fig.~\textcolor{darkteal}{S10}\ignore{\ref{fig_SI:rms}}), but Cs$_2$Te has a larger average atomic mass (Fig.~\textcolor{darkteal}{S7}\ignore{\ref{fig_SI:velo_ltc_avg_mass}}) and lower values of both $\langle \nu \rangle$ and $\kappa_{\mathrm{L}}$ (Fig.~\ref{fig:velo_ltc}). Figure~\textcolor{darkteal}{S14}\ignore{\ref{fig_SI:Cs2Se_Cs2Te}} compares the phonon dispersions, group velocities, and cumulative $\kappa_{\mathrm{L}}$ of the two compounds. Their acoustic and low-frequency optical modes predominantly involve Cs vibrations or mixed vibrations of Cs and the chalcogen atoms. The group velocities of these modes are comparable in the two compounds. The clearest difference occurs in the high-frequency optical modes, which predominantly involve chalcogen vibrations. The lighter Se atoms in Cs$_2$Se contribute to markedly higher group velocities for these modes.

Cs$_2$HgTe$_2$($I4/mmm$, No.~139) likewise has a cation--anion $\langle \mathrm{RMS}(\Phi_{ij}^{(2)}) \rangle$ close to that of Cs$_2$Se (Fig.~\textcolor{darkteal}{S10}\ignore{\ref{fig_SI:rms}}). Its larger average atomic mass (Fig.~\textcolor{darkteal}{S7}\ignore{\ref{fig_SI:velo_ltc_avg_mass}}) likely contributes to its lower $\langle \nu \rangle$ (Fig.~\ref{fig:velo_ltc}). Rb$_4$CdCl$_6$ further illustrates the combined effects of coupling strength and atomic mass. It has relatively large cation--anion force constants and the second-lowest average atomic mass in this category (Fig.~\textcolor{darkteal}{S7}\ignore{\ref{fig_SI:velo_ltc_avg_mass}}). Together, these factors favor phonon propagation and contribute to its relatively high $\langle \nu \rangle$.

\subsubsection{Heavy-element-filled host–guest-like structures}
Figure~\ref{fig:stru}(i--k) shows representative structures in this category. These compounds contain connected host frameworks that take the form of one-dimensional chains, two-dimensional layers, or cage-like networks. Heavy guest atoms occupy interlayer regions, channels, or cavities and commonly provide charge balance. The connected frameworks provide pathways for phonon propagation, consistent with the generally higher $\kappa_{\mathrm{L}}$ of these compounds than that of molecular crystals and structures with discrete anionic units (Fig.~\ref{fig:velo_ltc}). The heavy guest atoms, however, are typically weakly coupled to the host. As shown in Fig.~\textcolor{darkteal}{S15}\ignore{\ref{fig_SI:guest_rms}}, the $\langle \mathrm{RMS}(\Phi_{ij}^{(2)}) \rangle$ is below 0.25 eV/\AA$^{2}$ for most compounds in this category and reaches at most 0.41 eV/\AA$^{2}$. The weak restoring forces permit large-amplitude guest vibrations, contributing to large mean-square displacements. Together with the large guest-atom masses, these weak restoring forces give rise to low-frequency, guest-dominated optical modes and a small $\omega_{\min}^{\mathrm{opt}}$. These modes are often localized or weakly dispersive. Their coupling to heat-carrying acoustic phonons can flatten phonon branches and reduce group velocities~\cite{Euchner2012}, as well as enhance phonon scattering~\cite{PhysRevB.93.205203}. Weakly bound heavy guest atoms can therefore limit heat transport even in structures with connected host frameworks.

Cs$_3$As and Rb$_3$Bi deviate markedly from the trend between $\langle \nu \rangle$ and $\kappa_{\mathrm{L}}$ in this category. Their average phonon group velocities of 540 and 520 ms$^{-1}$, respectively, are close to that of K$_2$RbSb (520 ms$^{-1}$). Their $\kappa_{\mathrm{L}}$ values, however, are substantially lower: 0.41 Wm$^{-1}$K$^{-1}$ for Cs$_3$As and 0.38 Wm$^{-1}$K$^{-1}$ for Rb$_3$Bi, compared with 0.73 Wm$^{-1}$K$^{-1}$ for K$_2$RbSb. In K$_2$RbSb, Cs$_3$As, and Rb$_3$Bi, the connected host frameworks consist of K--Sb, Cs--As, and Rb--Bi, respectively, and the remaining alkali-metal atoms occupy framework voids (Fig.~\textcolor{darkteal}{S16}\ignore{\ref{fig_SI:Rb3Bi-Cs3As_stru}}). Figure~\textcolor{darkteal}{S17}\ignore{\ref{fig_SI:Rb3Bi-Cs3As}}(a--c) compares the phonon dispersions, three- and four-phonon scattering phase spaces, and scattering rates of these compounds. Both Cs$_3$As and Rb$_3$Bi exhibit substantially higher low-frequency scattering rates than K$_2$RbSb. Their larger three- and four-phonon scattering phase spaces allow more processes to satisfy energy and momentum conservation, contributing to the higher rates.

Comparison with Cs$_3$Bi also points to a role for anharmonicity. Cs$_3$Bi has a lower $\langle \nu \rangle$ of 440 ms$^{-1}$, yet its $\kappa_{\mathrm{L}}$ of 0.45 Wm$^{-1}$K$^{-1}$ exceeds that of either Cs$_3$As or Rb$_3$Bi. Its three- and four-phonon scattering phase spaces are comparable to those of the other two compounds, but its scattering rates are lower (Fig.~\textcolor{darkteal}{S17}\ignore{\ref{fig_SI:Rb3Bi-Cs3As}}(d)). Over the frequency range of approximately 0.3--1.0 THz, many modes in Cs$_3$As and Rb$_3$Bi have substantially larger absolute Gr\"uneisen parameters than those in Cs$_3$Bi (Fig.~\textcolor{darkteal}{S19}\ignore{\ref{fig_SI:Rb3Bi_Cs3As_gruneisen_compar}}). This difference indicates stronger low-frequency anharmonicity and is consistent with enhanced three-phonon scattering. Taken together, the comparisons with K$_2$RbSb and Cs$_3$Bi suggest that both the availability of multiphonon scattering channels and the strength of low-frequency anharmonic interactions contribute to the unusually low $\kappa_{\mathrm{L}}$ of Cs$_3$As and Rb$_3$Bi.

Cs$_2$SnTe$_2$ ($Ibam$, No.~72) and CsKAu$_2$ ($Fm\bar{3}m$, No.~225) exhibit particularly low $\kappa_{\mathrm{L}}$ values of 0.18 and 0.16 Wm$^{-1}$K$^{-1}$, respectively, at 300 K. Their structures are shown in Fig.~\textcolor{darkteal}{S20}\ignore{\ref{fig_SI:Cs2SnTe2}} and Fig.~\textcolor{darkteal}{S21}\ignore{\ref{fig_SI:CsKAu2}}, respectively. CsKAu$_2$ adopts the full-Heusler structure, a structural family for which ultralow $\kappa_{\mathrm{L}}$ has previously been predicted~\cite{PhysRevLett.117.046602}. To our knowledge, the lattice thermal conductivity of CsKAu$_2$ itself has not previously been investigated.

In both compounds, soft host frameworks contribute to low phonon group velocities. In Cs$_2$SnTe$_2$, the nearest-neighbor Sn--Te interaction has an $\mathrm{RMS}(\Phi_{ij}^{(2)})$ of 0.698 eV/\AA$^{2}$, indicating relatively weak coupling within the host framework. The Cs guest atoms are also weakly coupled to the host, with nearest-neighbor Cs--Te and Cs--Sn values of 0.16 and 0.20 eV/\AA$^{2}$, respectively. In CsKAu$_2$, the Cs--Au interactions within the host framework have an $\langle \mathrm{RMS}(\Phi_{ij}^{(2)}) \rangle$ of 0.201 eV/\AA$^{2}$. The corresponding K--Au value is approximately 0.1 eV/\AA$^{2}$, indicating still weaker coupling between the K guest atoms and the host. These weak interactions favor weakly dispersive phonon branches and low group velocities.

Strong phonon scattering further suppresses heat transport in both compounds, although the modes involved differ. In Cs$_2$SnTe$_2$, low-frequency, Te-dominated optical modes overlap and couple with acoustic modes, providing numerous scattering channels and contributing to a large scattering phase space (Fig.~\textcolor{darkteal}{S22}\ignore{\ref{fig_SI:Cs2SnTe2-CsKAu2}}(a)). CsKAu$_2$ has a smaller scattering phase space and lower scattering rates than Cs$_2$SnTe$_2$, although its phase space remains large and its scattering rates remain high. In CsKAu$_2$, enhanced scattering primarily involves modes in which both Cs and Au participate (Fig.~\textcolor{darkteal}{S22}\ignore{\ref{fig_SI:Cs2SnTe2-CsKAu2}}(b)). The $-\mathrm{pCOHP}$ analysis in Fig.~\textcolor{darkteal}{S23}\ignore{\ref{fig_SI:CsKAu2_Cs2SnTe2_cohp}} reveals pronounced occupied Cs--Au antibonding states, which may contribute to anharmonicity and enhanced scattering. Thus, low group velocities and high scattering rates jointly account for the exceptionally low $\kappa_{\mathrm{L}}$ of both compounds.

Some compounds with filled host frameworks exhibit $\langle \nu \rangle$ and $\kappa_{\mathrm{L}}$ values comparable to those of structures with discrete anionic units separated by heavy cations (Fig.~\ref{fig:velo_ltc}). One example is Cs$_2$HgPb ($Cmcm$, No.~63), with $\langle \nu \rangle$ = 300 ms$^{-1}$ and $\kappa_{\mathrm{L}}$ = 0.337 Wm$^{-1}$K$^{-1}$. Its host framework consists of one-dimensional Hg--Pb chains (Fig.~\textcolor{darkteal}{S24}\ignore{\ref{fig_SI:Cs2HgPb_stru}}). The Hg--Pb $\langle \mathrm{RMS}(\Phi_{ij}^{(2)}) \rangle$ reaches 1.29 eV/\AA$^{2}$, whereas the maximum $\mathrm{RMS}(\Phi_{ij}^{(2)})$ values for Cs--Hg and Cs--Pb are only 0.09 and 0.15 eV/\AA$^{2}$, respectively. These values indicate relatively strong interactions within the chains and weak coupling between the Cs guests and the host. Cs$_2$HgPb has the largest average atomic mass in this category (Fig.~\textcolor{darkteal}{S7}\ignore{\ref{fig_SI:velo_ltc_avg_mass}}), which favors low phonon frequencies and group velocities. Its phonon spectrum extends only to 4.07 THz, and its scattering rates are relatively low (Fig.~\textcolor{darkteal}{S25}\ignore{\ref{fig_SI:Cs2HgPb-Cs3InSb2}}(a)). These observations suggest that the low $\kappa_{\mathrm{L}}$ of Cs$_2$HgPb arises primarily from the low group velocities associated with its large average atomic mass.

Cs$_2$NaSb also exhibits low $\langle \nu \rangle$ and $\kappa_{\mathrm{L}}$, although its average atomic mass is unexceptional within this category (Fig.~\textcolor{darkteal}{S7}\ignore{\ref{fig_SI:velo_ltc_avg_mass}}). The $-\mathrm{ICOHP}$ values are 0.48 for Na--Sb and 0.23 for Cs--Sb, indicating stronger bonding between Na and Sb. The Na and Sb atoms form a connected two-dimensional network, with Cs atoms occupying sites within the framework (Fig.~\textcolor{darkteal}{S26}\ignore{\ref{fig_SI:Cs2NaSb}}). Both the connectivity and the bonding analysis thus identify the Na--Sb network as the host. Even within this network, the maximum Na--Sb $\mathrm{RMS}(\Phi_{ij}^{(2)})$ is only 0.16 eV/\AA$^{2}$, comparable to the weak cation--anion interactions in the second category. In Cs$_2$NaSb, weak interactions within the host framework therefore favor weakly dispersive phonon branches and low group velocities, contributing to its low $\kappa_{\mathrm{L}}$.

\subsubsection{Weakly bonded continuous-network structures}
Figure~\ref{fig:stru}(l) illustrates the continuous networks characteristic of this category. These networks lack the isolated molecular units, discrete anionic motifs separated by heavy cations, and distinct host--guest arrangements that characterize the preceding categories. All compounds identified in this category are binary, possibly because their simpler chemical environments favor a more uniform distribution of bond strengths. In compounds containing more than two elements, a broader distribution of bond strengths may give rise to distinct structural units or host frameworks. Weaker interactions can then connect these units or bind guest atoms to the framework. Such compounds are therefore more likely to fall into one of the preceding categories. Most members of the continuous-network category are alkali-metal halides, in which the large electronegativity difference between the alkali metal and the halogen favors ionic bonding.

Figure~\textcolor{darkteal}{S27}\ignore{\ref{fig_SI:wcb_rms}} shows $\langle \mathrm{RMS}(\Phi_{ij}^{(2)}) \rangle$ between cations and anions in this category. Among the alkali-metal halides, even the largest value, obtained for RbBr ($F\bar{4}3m$, No.~216), is only 0.31 eV/\AA$^{2}$. These small force constants indicate weak restoring forces for atomic displacements within the continuous networks. Such weak restoring forces favor large atomic mean-square displacements and, together with the large atomic masses, give rise to low-frequency optical modes and a small $\omega_{\min}^{\mathrm{opt}}$. The combination of a soft lattice and large atomic masses also favors weakly dispersive phonon branches and low $\langle \nu \rangle$, thereby limiting heat transport. Thermal transport in alkali-metal halides and related binary compounds has been studied extensively~\cite{10.1002/adfm.202108532,PhysRevX.10.041029,chang2022,10.1021/acsaelm.3c00759,PhysRevB.110.035205}.

The only members of this category outside the alkali-metal halide family are AgCl ($F\bar{4}3m$, No.~216), CuCl ($F\bar{4}3m$, No.~216), and Ag$_2$S ($Fd\bar{3}m$, No.~227). Their cation--anion $\langle \mathrm{RMS}(\Phi_{ij}^{(2)}) \rangle$ values are substantially larger, reaching 0.69, 0.81, and 2.72 eV/\AA$^{2}$, respectively (Fig.~\textcolor{darkteal}{S27}\ignore{\ref{fig_SI:wcb_rms}}). Strong acoustic phonon scattering is the primary contributor to the low $\kappa_{\mathrm{L}}$ of these three compounds (Fig.~\textcolor{darkteal}{S28}\ignore{\ref{fig_SI:AgCl}}). Four-phonon scattering is particularly pronounced in CuCl and AgCl, whereas both three- and four-phonon processes contribute substantially in Ag$_2$S.

The electronic bonding in these compounds provides a qualitative explanation for the high scattering rates. Previous studies have linked hybridization between Cu or Ag $d$ orbitals and anion $p$ orbitals to the formation of occupied antibonding states. Occupation of these states can increase the anharmonicity of the potential energy associated with specific atomic displacements~\cite{10.1002/adfm.202108532,https://doi.org/10.1002/advs.202417292}, due to weakened Cu/Ag-anion chemical bonds. The $-\mathrm{pCOHP}$ curves in Fig.~\textcolor{darkteal}{S29}\ignore{\ref{fig_SI:AgCl_cohp}} show pronounced antibonding character below the Fermi level in all three compounds. Occupation of these antibonding states can weaken the corresponding bonds and may contribute to the anharmonic interactions responsible for the strong phonon--phonon scattering.

\section{Conclusions}
Although lattice thermal conductivity depends on anharmonic phonon interactions, which are more computationally expensive to calculate, the harmonic vibrational spectrum contains signatures of exceptionally weak or efficient heat transport. In this work, we show that two quantities derived from harmonic vibration, the average atomic mean-square displacement and the lowest zone-center optical phonon frequency, can guide the search for semiconductors at both conductivity extremes. Across the evaluated compounds from \textsc{Phonix} database, larger atomic mean-square displacement correlated with lower Debye sound velocity, while lower lowest-zone-center optical phonon frequency correlated with a larger weighted phase space for three-phonon scattering. The descriptors thus connect two quantities accessible in a harmonic calculation to two distinct aspects of heat transport: how rapidly acoustic phonons propagate and which three-phonon scattering processes are available to them.

To resolve these signatures at database scale, we fine-tuned \textsc{MatterSim} using second-order interatomic force constants from \textsc{PhononDB} as the primary training target and atomic forces as an auxiliary target. Training on the curvature of the potential-energy surface reduced the mean absolute error in phonon frequencies from 0.153 to 0.109~THz and increased $R^2$ from 0.808 to 0.923 for atomic mean-square displacement and from 0.935 to 0.953 for the lowest zone-center optical phonon frequency.

Of the 8,974 compounds from the Open Quantum Materials Database that passed harmonic phonon screening based on fine-tuning \textsc{MatterSim}, the two descriptors selected 463 low-$\kappa_{\mathrm{L}}$ candidates and 49 high-$\kappa_{\mathrm{L}}$ candidates. First-principles transport calculations at 300~K predicted $\kappa_{\mathrm{L}}$ < 1 Wm$^{-1}$K$^{-1}$ for 104 of the 149 low-conductivity candidates evaluated and $\kappa_{\mathrm{L}}$ > 100 Wm$^{-1}$K$^{-1}$ for 15 of the 34 high-conductivity candidates evaluated. The recovery of established high-conductivity semiconductors, including BAs, BN, and SiC, provides a complementary test of the same physical criteria used to identify ultralow-conductivity compounds.

The 104 compounds predicted to exhibit ultralow $\kappa_{\mathrm{L}}$ span molecular crystals and related structures, crystals with discrete anionic units separated by heavy cations, host--guest structures, and weakly bonded continuous networks. Across these families, the average phonon group velocity correlated more strongly with $\kappa_{\mathrm{L}}$ than did the Debye sound velocity. Thus, while long-wavelength acoustic softening helps identify candidates, phonon propagation across the Brillouin zone better distinguishes conductivities within the ultralow-$\kappa_{\mathrm{L}}$ regime. The soft frameworks of $\mathrm{CsKAu_2}$ and $\mathrm{Cs_2SnTe_2}$ illustrate how strong anharmonic scattering can further suppress heat transport when phonon velocities are already low, yielding predicted room-temperature conductivities of 0.16 and $0.18\,\mathrm{W\,m^{-1}\,K^{-1}}$, respectively.

These findings provide a common physical basis for searching across crystal families whose bonding and connectivity differ substantially. Because the screening criteria are defined by lattice dynamics, they can be carried over to other interatomic potentials as their accuracy for soft vibrational modes improves. This offers a practical route to directing anharmonic calculations toward materials of interest for thermal insulation, thermoelectric energy conversion, and heat dissipation.

\section{ACKNOWLEDGMENTS}
The authors acknowledge the support received from the National Natural Science Foundation of China (Grant No.\,12374024), Fundamental Research Funds for the Central Universities (No. FRF-BRB-25-006), and State Key Laboratory for Advanced Metals and Materials, Grant No. 2025-Z15.

\FloatBarrier
\bibliographystyle{elsarticle-num}
\bibliography{refs}



\end{document}